\documentclass[twocolumn,floatfix,prb,aps,showpacs]{revtex4-1}
\usepackage{graphicx,epsf,amsmath,amssymb}
\usepackage{bm,bbm}
\usepackage{multirow}
\usepackage{import}
\usepackage{enumerate}
\usepackage{hyperref}
\hypersetup{
    bookmarks=true,         
    unicode=false,          
    pdftoolbar=true,        
    pdfmenubar=true,        
    pdffitwindow=false,     
    pdfstartview={FitH},    
    pdftitle={Surface states of Weyl semimetals -- Serguei Tchoumakov --  LPS},    
    pdfauthor={S.Tchoumakov},     
    pdfsubject={Report},   
    pdfcreator={S.Tchoumakov},   
    pdfproducer={S.Tchoumakov}, 
    pdfkeywords= {electrons} {weyl} {surface} 
    pdfnewwindow=true,      
    colorlinks=true,       
    linkcolor=blue,          
    citecolor=red,        
    filecolor=magenta,      
    urlcolor=cyan           
}

\usepackage[usenames,dvipsnames,svgnames,table]{xcolor}

\newcommand{\executeiffilenewer}[3]{%
    \ifnum\pdfstrcmp{\pdffilemoddate{#1}}%
        {\pdffilemoddate{#2}}>0{\immediate\write18{#3}}
    \fi}

\begin{document}
\title{Magnetic description of the Fermi arc in type-I and type-II Weyl semimetals} 

\author{Serguei Tchoumakov, Marcello Civelli and Mark O. Goerbig$^1$}
\affiliation{$^{1}$Laboratoire de Physique des Solides, Univ. Paris-Sud, Universit\'e Paris-Saclay, CNRS UMR 8502, F-91405 Orsay Cedex, France}
\date{\today}

\begin{abstract}
We consider finite-sized interfaces of a Weyl semi-metal and show that the corresponding confinement potential is similar to the application of a magnetic field. Among the numerous states, which can be labeled by indices $n$ like in Landau levels, the $n = 0$ surface state describes the Weyl semimetal Fermi arc at a given chemical potential. Moreover, the analogy with a magnetic field shows that an external in-plane magnetic field can be used to distort the Fermi arc and would explain some features of magneto-transport in Weyl semimetals. We derive the Fermi arc for type-I and type-II Weyl semimetals where we deal with the tilt anisotropy by the use of Lorentz boosts. In the case of type-II Weyl semimetals, this leads to many additional topologically trivial surface states at low energy. Finally, we extend the Aharonov-Casher argument and demonstrate the stability of the Fermi arc over fluctuations of the surface potential. 
\end{abstract}


\maketitle
\section{Introduction} 
One key feature of the interface of a topological insulator put into contact with a normal insulator is the presence of stable surface states [\onlinecite{reviewtopo}]. This stability is related to the topological Chern number which is based on the existence of a band gap. Surface states can play a crucial role since their metallicity can radically affect the transport properties of the topological insulator.  There are however topological materials, Dirac semimetals (DSM) and Weyl semimetals (WSM), where the band gap is exactly zero and they can be described as a critical phase lying between normal and topological insulators [\onlinecite{reviewwsm},\onlinecite{arpeswsmrev}].

In recent measurements [\onlinecite{wsmss0}-\onlinecite{wsmss2}] on materials of the (Ta,Nb)-(As,P) family of three-dimensional (3D) WSM, surface states with surprising properties were observed and related to the Fermi arc, as described in numerous theoretical studies [\onlinecite{wsmsstheo1,wsmsstheo2, wsmsstheo3,wsmsstheo4,wsmssmag1}]. In particular, the Fermi arc surface states have an open Fermi surface that connects the projected bulk Weyl nodes and, furthermore, they are spin polarized. These properties are used to characterize genuine WSM [\onlinecite{arpeswsmrev}, \onlinecite{wsmss1}]. Therefore, a better understanding of the Fermi arc properties has been sought in many theoretical investigations. Recently similarities between Fermi arcs and Landau bands [\onlinecite{wsmssmag1},\onlinecite{wsmssmag2}] have been pointed out, although the latter typically arise in the presence of a magnetic field.

In various theoretical studies, the crystal potential at the surface of the WSM is sharp because it simplifies the derivation of the wavefunction boundary conditions [\onlinecite{wsmsstheo1,wsmsstheo2, wsmsstheo3,wsmsstheo4}]. In this approach one considers that the width of the interface $\ell$ is shorter than any other length scale, such as the magnetic or the band bending lengths. In some situations, however, this description is not sufficient. For example, in Ref. [\onlinecite{majoranass}], in order to describe the effect of a magnetic domain wall on Majorana bound states of a superconducting chain, it was necessary to introduce a smooth gap inversion at the interface. 

\begin{figure}[thb]
    \centering
    \includegraphics[width=\columnwidth]{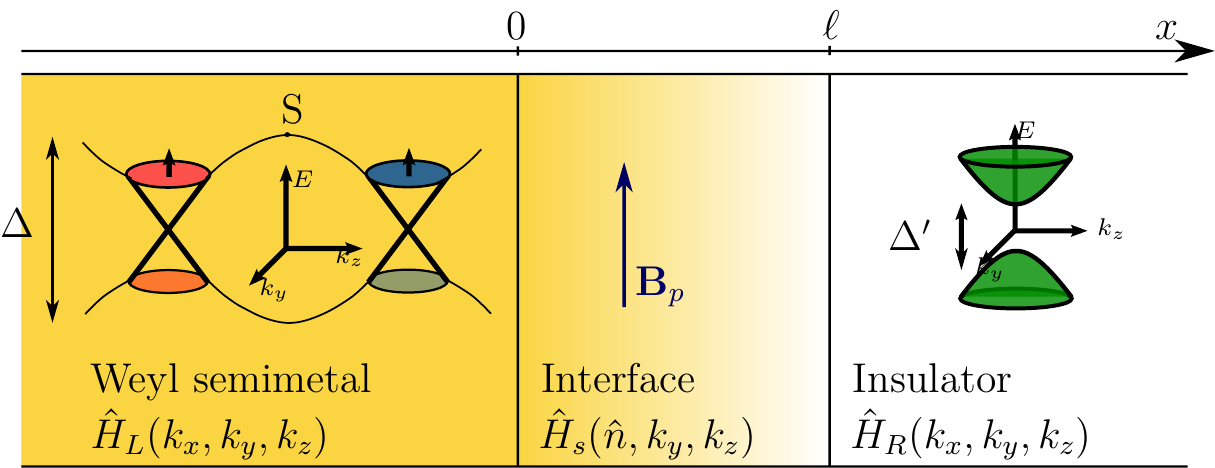}
    \caption{(Color online) Sketch of the smoothly varying interface between two bulk materials, a Weyl semimetal with two Weyl cones separated by a saddle point at energy $\Delta$
(in yellow, on the left) and an insulator with gap $\Delta'$ (in white, on the right). Within the interface of width $\ell$, the electrons behave as if they were immersed into an in-plane magnetic field ${\bf B}_p$. }
    \label{fig:interface_general}
\end{figure}

In the present paper, we extend the description of a smoothly varying interface to the WSM in order to describe the coupling to external electromagnetic fields. We consider that two bulk materials are put into contact with an interface of width $\ell$ (see Fig. {\ref{fig:interface_general}). 
In agreement with recent semiclassical [\onlinecite{wsmssmag1}] and numerical [\onlinecite{wsmssmag2}] studies, we show, within a constructive quantum-mechanical approach, that the Hamiltonian describing the interface is similar to one in the presence of a magnetic field within a particular gauge. This leads to a set of two-dimensional (2D) Landau bands indexed by an integer number $n$. In the case of WSM, these states have open Fermi surfaces as a consequence of the localization of Landau orbits inside the region of size $\ell$, which is similar to the derivation of the degeneracy of Landau bands within the Landau gauge [\onlinecite{markreview}]. This yields a Fermi surface with an odd number of Fermi arcs with only the $n = 0$ Landau band having a single Fermi arc. This state is the most relevant at low energy and is the main subject of our investigation. The analogy with the magnetic field allows us to explore various aspects of the $n = 0$ Fermi arc. In particular, we show that the effect of an external magnetic field is to modify the shape of the Fermi arc and that the combination of tilt and magnetic field can break the arc into two separate pieces. This is investigated in the case of type-I and type-II WSM [\onlinecite{wsmtype2}]. 
Moreover, in spite of the reduced dimension of the surface compared to the bulk, we show that the transport properties are widely affected by the surface states. Indeed, the $n = 0$ surface state acts as a Faraday cage and localizes the electric field at the surface.
We finally prove the topological stability of the $n = 0$ Fermi arc, based on a generalization of the Aharonov-Casher argument [\onlinecite{acasher}].

The paper is organized as follows: in Sec. \ref{sec:intefmodel} we explicitly model the interface of a bulk Weyl semimetal with an insulating bulk. Section \ref{sec:apscreening} is devoted to a detailed discussion of the surface and bulk screening. In Sec. \ref{sec:electricregime} we describe the effect of the tilt anisotropy on the Fermi arc structure and relate it to the effect of an electric field which is canceled with a proper use of Lorentz boosts. 
We extend the Aharonov-Casher argument in Sec. \ref{sec:topostability}, originally established to prove the stability of the $n = 0$ Landau level of 2D DSM in a spatially inhomogeneous magnetic field, to the surface states of 3D WSM in the presence of a fluctuating surface potential. We extend the use of chirality to three-dimensional systems and relate the long range behavior to surface states derived using boundary conditions. We present a detailed discussion of our results in Sec. \ref{sec:discussion}, discussing possible consequences on magneto-transport experiments.

\section{Magnetic description of the surface states}\label{sec:intefmodel}
 
Let us first consider a general case of a left and a right bulk, each of which is modeled by a set of
parameters (see Fig.~\ref{fig:interface_general}). These parameters vary smoothly across an interface of
size $\ell$, and here we choose to linearly interpolate them between the two bulk Hamiltonians: $\hat{H}_L = {\bf h}_L({\bf k}) \cdot \hat{\boldsymbol{\sigma}}$ in the material on the left side ($x < 0$) and $\hat{H}_R = {\bf h}_R({\bf k}) \cdot \hat{\boldsymbol{\sigma}}$ in the material on the right side ($x > \ell$), where $\hat{\boldsymbol{\sigma}} = (\hat{\sigma}_x, \hat{\sigma}_y, \hat{\sigma}_z)$ is a vector combining the three Pauli matrices. The interface region $x \in [0,\ell]$ between the two bulk materials can also model a surface in the case where one of the two materials is a trivial bulk insulator.
The discussion of the general model in Sec.~\ref{ssec:interface} allows us to investigate the emergence of Fermi arcs on the surface, while we discuss more explicit examples in the following two subsections. In Sec.~\ref{ssec:merging}, we discuss a pair of Weyl points that merge in the interface region opening a gap. Finally in Sec.~\ref{sec:fractiont0} we analyze a simplified model of single Weyl cones.

\subsection{Interface Hamiltonian}
\label{ssec:interface}

We consider as Hamiltonian $\hat{H}_s$ describing the region of the interface, $x\in[0,\ell]$, the linear interpolation between the two bulk Hamiltonians $\hat{H}_L$ and $\hat{H}_R$
\begin{align}\label{eq:interfH}
	\hat{H}_s = \left[ {\bf h}_L + \delta \mathbf{h} x/\ell \right] \cdot \hat{\boldsymbol{\sigma}}
\end{align}
where we introduce $\delta \mathbf{h}({\bf k}) = \mathbf{h}_R({\bf k}) - \mathbf{h}_L({\bf k})$, such that $\hat{H}_s(x = 0) = \hat{H}_L$ and $\hat{H}_s(x=\ell) = \hat{H}_R$. We discuss the validity of this linear interpolation in Sec.~\ref{sec:topostability}, where we show that deviations in the form of a fluctuating interface potential do not alter the main conclusions drawn from the present model.
Notice that the wavevector component $k_x$ perpendicular to the interface needs to be treated as a quantum-mechanical operator such that $[x,k_x]=i$ (here, and in the remaining parts of this paper, we use a system of units with $\hbar=1$). We consider situations with a linear $x$-component in the Hamiltonian that is unchanged across the interface, such that $\hat{H}_{L}(k_x) - \hat{H}_{L}(k_x = 0) = \hat{H}_{R}(k_x) - \hat{H}_{R}(k_x = 0) = v_x k_x \hat{\sigma}_x$ and $\delta {\bf h} = \delta {\bf h}({\bf k}_{\parallel})$ with the inplane momentum ${\bf k}_{\parallel} = (0, k_y, k_z)$. 
In order to describe the opposite interface one has to permute ${\bf h}_L \leftrightarrow {\bf h}_R$. This corresponds to replacing $({\bf h}_L, \delta {\bf h}, \ell)$ in Eq.~(\ref{eq:interfH}) by $({\bf h}_R, -\delta {\bf h}, \ell)$ or equivalently by $({\bf h}_R, \delta {\bf h}, -\ell)$ in all the derived expressions.

We rotate the Hamiltonian $\hat{H}_s$ with angle $\theta$ along the $x-$axis with $\tan(\theta) = \delta h_z/\delta h_y$ in order to simplify the treatment of the non-commuting  variables, $[x,k_x] \neq 0$. In the new basis, the interface Hamiltonian reads
\begin{align}\label{eq:rotH}\nonumber
	\hat{H}_s^{(\theta)} &= e^{i \theta \hat{\sigma}_x/2}\hat{H}_se^{-i \theta \hat{\sigma}_x/2}\\
	&= v_x \left[ \left(  k_x + x/\ell_x^2 \right) \hat{\sigma}_x + {\rm sign}(v_x \ell)\frac{x - \langle x \rangle}{\ell_S^2} \hat{\sigma}_y \right]\nonumber\\
	&~~~~~~ + M({\bf k}_{\parallel}) \hat{\sigma}_z,
\end{align}
and we introduce the notation $\delta {\bf h}_{\parallel}({\bf k}_{\parallel}) = (0, \delta h_y, \delta h_z)$ for the in-plane components of $\delta {\bf h}$ and
\begin{align}\label{eq:posdef}
		\begin{array}{l}
			\langle x \rangle/\ell = - \delta {\bf h}_{\parallel}\cdot {\bf h}_L/{\delta {\bf h}_{\parallel}^2} = 1 - \delta {\bf h}_{\parallel}\cdot {\bf h}_R/{\delta {\bf h}_{\parallel}^2},\\
			\ell_x = \sqrt{|v_x \ell|/|\delta h_x|},~~~~~ \ell_S = \sqrt{|v_x \ell|/|\delta {\bf h}_{\parallel}|},\\
			M({\bf k}_{\parallel}) = {\bf b}\cdot {\bf h}_L/|{\bf b}| = {\bf b}\cdot {\bf h}_R/|{\bf b}|,
		\end{array}
\end{align}
where $\mathbf{b} = \boldsymbol\nabla \times (\delta \mathbf{h} x/\ell) = \mathbf{e}_x \times \delta {\bf h}/\ell$ plays a role similar to the magnetic field as we show below. 
Hamiltonian (\ref{eq:rotH}) can then be rewritten with the help of the following ladder operators
\begin{align}\label{eq:ladders}
	\begin{array}{l}
		\hat{a} = \frac{\ell_S}{\sqrt{2}}\left[ k_x + x/\ell_x^2 - i~{\rm sign}(v_x \ell) (x - \langle x \rangle)/{\ell_S^2}\right],\\~~\\
		\hat{a}^{\dagger} = \frac{\ell_S}{\sqrt{2}}\left[ k_x + x/\ell_x^2 + i~{\rm sign}(v_x \ell) (x - \langle x \rangle)/\ell_S^2\right],
	\end{array}
\end{align}
such that $\left[ \hat{a}, \hat{a}^{\dagger} \right] = {\rm sign}(v_x \ell)$, and the interface Hamiltonian in the rotated basis thus becomes
\begin{align}\label{eq:Htransfofinale}
	\hat{H}_s^{(\theta)} &=
	\left( 
		\begin{array}{cc}
			M({\bf k}_{\parallel}) & \sqrt{2} |v_x| \hat{a}/\ell_S\\
			\sqrt{2} |v_x| \hat{a}^{\dagger}/\ell_S & - M({\bf k}_{\parallel})
		\end{array}
	\right).
\end{align}
In the case $v_x \ell > 0$, the eigenstates can be written in the form
\begin{align}
	|\Psi\rangle = \left(
		\begin{array}{c}
			\alpha_n |n-1\rangle\\
			\beta_n |n\rangle
		\end{array}
	\right)
\end{align}
where $|n\rangle$ are the eigenstates of the number operator $\hat{n} = \hat{a}^{\dagger}\hat{a}$ of eigenvalue $n$. In the other case where $v_x \ell < 0$, the spinor components are interchanged. These states are located around $\langle x \rangle = - \delta {\bf h}_{\parallel}\cdot {\bf h}\ell/{\delta {\bf h}_{\parallel}^2}  \sim \ell$, oscillating in the $x-$direction with a wavelength $\lambda_x = 2\pi/\langle k_x \rangle = 2\pi \ell_x^2/\langle x \rangle \sim \ell$, and they spread over the typical size $\Delta x = \sqrt{2n}\ell_S \sim \sqrt{\ell}$.

We find that the $n\geq 1$ bands are gapped with eigenstates in the original basis $|\Psi_{\lambda,n}\rangle$ and eigenenergies $E_{\lambda,n}$ ($\lambda = \pm 1$) such that 
\begin{align}\label{eq:solrot}
		\begin{array}{l}
			|\Psi_{\lambda,n}\rangle = 
			e^{i \theta \hat{\sigma}_x /2} \left( 
				\begin{array}{c}
					\left[ 1 + \lambda \frac{M({\bf k}_{\parallel})}{E_n} \right] | n - 1 \rangle \\
					\lambda \left[ 1 - \lambda \frac{M({\bf k}_{\parallel})}{E_n} \right]|n\rangle
				\end{array}
			\right),\\~~\\
		E_{\lambda,n}(\mathbf{k}_{\parallel}) = \lambda E_n = \lambda \sqrt{M({\bf k}_{\parallel})^2 + 2 v_x^2 n/\ell_S^2}.
		\end{array}
\end{align}
These solutions are similar to Landau bands of a Weyl semimetal in a pseudo-magnetic field $\mathbf{B}_p$ along ${\bf b}$ and of magnetic length $\ell_S$ \cite{}. The gaps between the bands are on the order of $\sqrt{2}|v_x|/\ell_S \sim 1/\sqrt{\ell} \rightarrow \infty$, i.e. they diverge
with $\ell \rightarrow 0$ and are thus usually neglected in the discussion of sharp interfaces [\onlinecite{wsmsstheo1}]. In the case $\ell \neq 0$, the $n \geq 1$ states should be observed at higher energies and in section \ref{sec:electricregime} we show that a tilt in the band structure can reduce their gaps.

The $n = 0$ Landau band which we are interested in here is special, in that it survives in the limit of sharp interfaces ($\ell \rightarrow 0$), and one finds either 
\begin{align}\label{eq:LLzerostate0}
		\begin{array}{l}
			|\Psi_{0}^+\rangle = 
			e^{i \theta \hat{\sigma}_x /2} \left( 
				\begin{array}{c}
					0 \\
					|0\rangle
				\end{array}
			\right),\\~~\\
		E_{0}^+(\mathbf{k}_{\parallel}) = -M({\bf k}_{\parallel}),
		\end{array}
\end{align}
for $v_x \ell > 0$, or
\begin{align}\label{eq:LLzerostate0bis}
		\begin{array}{l}
			|\Psi_{0}^-\rangle = 
			e^{i \theta \hat{\sigma}_x /2} \left( 
				\begin{array}{c}
					|0\rangle \\ 0
				\end{array}
			\right),\\~~\\
		E_{0}^-(\mathbf{k}_{\parallel}) = M({\bf k}_{\parallel}),
		\end{array}
\end{align}
for $v_x\ell < 0$. If one takes into account the rotation of the Hamiltonian in Eq.~(\ref{eq:rotH}), one finds that the isospin points along ${\bf e}_0 = - {\rm sign}(v_x\ell) \left( \delta h_{y}{\bf e}_z + \delta h_{z} {\bf e}_y \right)/\sqrt{\delta h_y^2 + \delta h_z^2}$, and the eigenenergy can be rewritten in a concise manner as $E_{0}({\bf k}_{\parallel}) = -{\rm sign}(v_x\ell)M({\bf k}_{\parallel})$.
Written explicitly, the $n=0$ surface states are given by the wave functions (apart from a normalization constant)
\begin{align}\label{eq:wavefon}
\nonumber
\psi_0^+(x,y,z) &\propto& \left(\begin{array}{c} \cos\frac{\theta}{2} \\ i\sin\frac{\theta}{2}
\end{array}\right)e^{-(x-\langle x\rangle)^2/2\ell_S^2}e^{-ix^2/2\ell_x^2}e^{i(k_yy + k_zz)}, \\
\nonumber
\psi_0^-(x,y,z) &\propto& \left(\begin{array}{c} i\sin\frac{\theta}{2} \\ \cos\frac{\theta}{2}
\end{array}\right)e^{-(x-\langle x\rangle)^2/2\ell_S^2}e^{-ix^2/2\ell_x^2}e^{i(k_yy + k_zz)},\\
\end{align}
where the sign in the superscript is ${\rm sign}(v_x\ell)$. The wavefunctions consist therefore of the usual harmonic-oscillator wavefunctions confined in the $x$-direction, centered around $\langle x\rangle$, multiplied by plane-wave functions in the $y$- and $z$-directions. Furthermore, one notices an additional phase $\exp(-ix^2/2\ell_x^2)$ that is due to the $x$-dependence of the real part of the harmonic-oscillator ladder operators (\ref{eq:ladders}).

The surface states must be localized in the interfacial region. As in the treatment of the degeneracy of Landau bands [\onlinecite{markreview}], this condition implies that the mean position $\langle x \rangle$ of the surface states is such that $\langle x \rangle \in [0,\ell]$. In section \ref{sec:topostability} we show that this same condition holds for the $n = 0$ surface state using another argument. This leads to the following localization condition 
\begin{align}\label{eq:consistencygeneral}
	- 1 < \delta {\bf h}_{\parallel}\cdot {\bf h}_L/{\delta {\bf h}_{\parallel}^2} < 0.
\end{align}
This condition is of uttermost importance in the understanding of the Fermi arc structure. In the following section, we illustrate this condition for the surface states of a Weyl semimetal.

\subsection{Fermi arc, merging of Weyl nodes}\label{ssec:merging}

After these general considerations, we now treat the situation with two Weyl nodes in the material on the left ($x < 0$) that merge at the interface and lead to an insulating state for $x > \ell$. This can be taken into account within the following merging model
\begin{align}\label{eq:hamqco}
	\hat{H}_{4} = \left( 
	\begin{array}{cc}
		v_F \left( \mathbf{k} + \Delta k_0 ~\mathbf{e}_z/2 \right) \cdot \hat{\sigma} & \kappa \mathbbm{1} \\
		\kappa \mathbbm{1} & -v_F \left( \mathbf{k} - \Delta k_0 ~\mathbf{e}_z/2 \right) \cdot \hat{\sigma}
	\end{array}	 \right),
\end{align}
where two Weyl nodes of opposite chirality are located at $k_z = \pm \Delta k/2$ with 
\begin{equation}\label{eq:dk}
\Delta k = \Delta k_0 \sqrt{1 - \left( \frac{2 \kappa}{v_F \Delta k_0}\right)^2}. 
\end{equation}
The two nodes merge at $\kappa = v_F \Delta k_0/2$ and then form a gapped phase for $\kappa > v_F\Delta k_0/2$.
In Appendix \ref{ap:fourbandreduction} we show that in the limit of strong coupling $|\kappa + \frac{v_F \Delta k_0}{2}| \gg |\kappa - \frac{v_F \Delta k_0}{2}|$, i.e. precisely in the vicinity of the merging transition we are interested in, this model can be reduced to the following two-band Hamiltonian
\begin{align}\label{eq:Hm}
	\hat{H}_{2L} = v_F (k_x \hat{\sigma}_x + k_y \hat{\sigma}_y ) + (k_z^2/2m - \Delta) \hat{\sigma}_z,
\end{align}
with $m = (\kappa + v_F \Delta k_0/2)/2 v_F^2$. One obtains either two Weyl nodes located at $k_z = \pm \Delta k/2 = \pm \sqrt{2m \Delta}$ if $\Delta > 0$ or a gapped phase if $\Delta < 0$, such that $\Delta$ now triggers the merging transition. Then the two bulk Hamiltonians are $\hat{H}_{2L} = v_F (k_x \hat{\sigma}_x + k_y \hat{\sigma}_y ) + (k_z^2/2m - \Delta) \hat{\sigma}_z$, $\Delta > 0$ for $x < 0$ and $\hat{H}_{2R} = v_F (k_x \hat{\sigma}_x + k_y \hat{\sigma}_y ) + (k_z^2/2m + \Delta') \hat{\sigma}_z$, $\Delta' > 0$ for $x > \ell$. We represent the two models by dashed lines in Fig.~\ref{fig:merging}, where we consider a continuous variation of the parameter $\delta h_z$, which is indeed given by $\Delta+\Delta'$, as we discuss below. On the left side, one has the phase with two Weyl nodes separated by $\Delta k$, and on the right side one finds the insulating state with gap $\Delta'$. Notice that, for $k_x = 0$ (or $k_y = 0$), this model reduces to that used in the study of Dirac point merging in 2D [\onlinecite{merging}]. 
\begin{figure}
    \centering
    \includegraphics[width=\columnwidth]{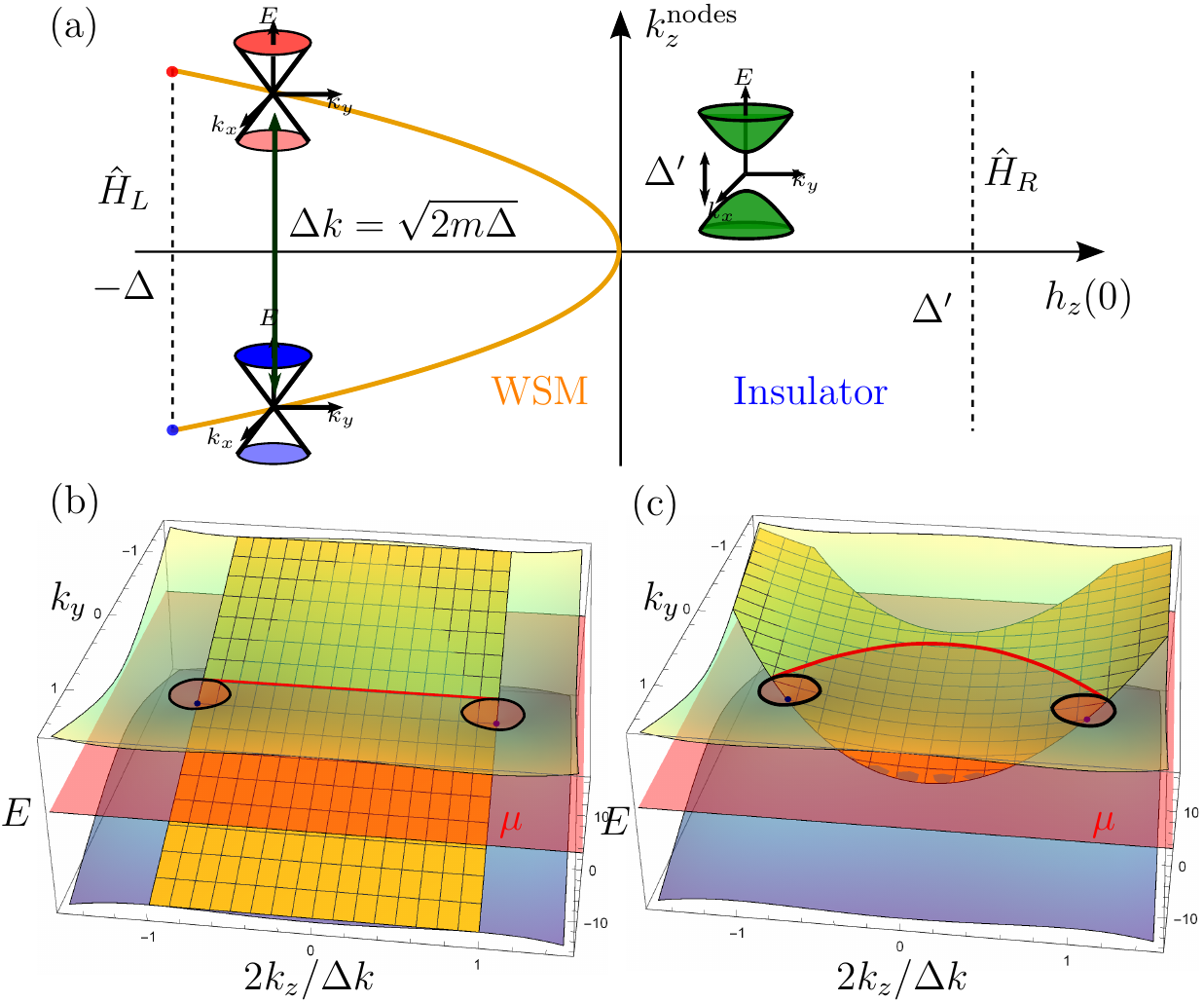}
    \caption{(Color online) (a) Evolution of the phase of the model in Eq.~(\ref{eq:Hsdoub}) for various values of $\delta h_z$, which varies linearly in $x$ in the interface. One finds an insulator for $\delta h_z > 0$ and a Weyl semimetal for $\delta h_z < 0$. The orange line indicates the location of the Weyl nodes along the $z$-axis. (b) and (c) Superposition of the bulk WSM band structure at $k_x = 0$ (transparent green for the conduction and transparent blue for the valence band), the $n = 0$ surface-states band structure for states that fulfill $\langle x \rangle \in [0,\ell]$ (in orange) and the plane of constant energy $\mu$ (chosen in the conduction band, in transparent red). The section at $k_x = 0$ of the bulk Fermi surface is represented by the thick black circles and the Fermi arc connecting the bulk Fermi surfaces by the thick red line. In (b) we consider no in-plane magnetic field $B \ll B_p$ (large gap insulator $\Delta'$) and recover the Fermi arc band structure found in other studies. In (c) the Fermi arc becomes curved for strong magnetic fields $B \sim B_p$  (small gap insulator $\Delta'$). 
}
    \label{fig:merging}
\end{figure}

We now assume that the parameter $\Delta$ varies linearly over the interface. 
In the notations introduced in the previous section, the bulk models are described by the following components
\begin{align}\label{eq:paramstwo}
		\begin{array}{l}
			{\bf h}_{2L}({\bf k}) = \left( v_F k_x, v_F k_y, k_z^2/(2m) - \Delta \right),\\
			\\
			{\bf h}_{2R}({\bf k}) = \left( v_F k_x, v_F k_y, k_z^2/(2m) + \Delta' \right),
		\end{array}
\end{align}
and, for $x \in [0,\ell]$, the interface Hamiltonian is
\begin{align}\label{eq:Hsdoub}
	\hat{H}_{2s} = v_F (k_x \hat{\sigma}_x& + k_y \hat{\sigma}_y )\nonumber\\
	 &+ \left(k_z^2/2m - \Delta + \frac{\Delta + \Delta'}{\ell}x\right) \hat{\sigma}_z,
\end{align}
i.e. $\delta {\bf h}$ reduces to $\delta {\bf h}=(0,0,\Delta + \Delta')$.
At this stage we can already introduce the effect of an external magnetic field, ${\bf B} = B {\bf e}_z$, along the direction separating the two cones. In the Landau gauge, the Peierls substitution leads to $k_y \rightarrow k_y + {\rm sign}(B)x/\ell_B^2$ with $\ell_B = 1/\sqrt{eB}$ the magnetic length. This is similar to the introduction of $\delta h_y \hat{\sigma}_y= {\rm sign}(B)v_F \ell/\ell_B^2\hat{\sigma}_y$ in Eq.~(\ref{eq:Hsdoub}), which corresponds to a displacement of the Weyl nodes over the interface from $k_y = 0$ to $k_y = \delta k_y \equiv -{\rm sign}(B) \ell/\ell_B^2$. This displacement can be introduced right from the beginning with a shift of $h_{2R,y}$ by a constant $\delta h_y^{0}$ and the same could be done with respect to $h_{2R,x}$. 
We consider a magnetic field along $z$ only for two reasons: (i) a component along $x$ would quantize the surface states, and (ii) for $\kappa \neq 0$ in Eq.~(\ref{eq:hamqco}), a component along $y$ is not along the principal axis of symmetry and thus difficult to solve, we only treat it in the case $\kappa = 0$ in section \ref{sec:fractiont0}.

As in the general case (\ref{eq:rotH}), we perform a rotation of angle $\theta$ of Hamiltonian (\ref{eq:Hsdoub}) along the $x-$axis and obtain a Hamiltonian similar to Eq.~(\ref{eq:Htransfofinale}),
\begin{align}\label{eq:HdoubleBp}
	\hat{H}^{(\theta)}_{2s} &= e^{i\frac{\theta}{2} \hat{\sigma}_x}\hat{H}_{2s} e^{-i\frac{\theta}{2} \hat{\sigma}_x}\\
	&= \left(
	\begin{array}{cc}
		M(\mathbf{k}_{\parallel}) & \sqrt{2}v_F \hat{a}_{2}/\ell_S\\
		\sqrt{2} v_F \hat{a}_{2}^{\dagger}/\ell_S & -M(\mathbf{k}_{\parallel})
	\end{array}
	\right),
\end{align}
for $\tan(\theta) = \ell_B^2/\ell_{B_p}^2$ where $\ell_{B_p} = 1/\sqrt{eB_p}$ is the magnetic length associated to the pseudo-magnetic field $B_p = (\Delta + \Delta')/(ev_F \ell)$ and $\ell_{S} = 1/\sqrt{eB_T}$ is the magnetic length associated to the total (effective) magnetic field $B_T = \sqrt{B_p^2 + B^2}$, combining both $B_p$ and $B$. 
In the same manner as in Eq.(\ref{eq:posdef})
we find the mass term $M(\mathbf{k}_{\parallel})= \left[B (k_z^2/2m-\Delta) - B_p v_F k_y\right]/B_T$ and the ladder operators $\hat{a}_2$,$\hat{a}_2^{\dagger}$, which define the number states $|n\rangle$ of mean position
\begin{align}\label{eq:definitionsFUS}
	&\frac{\langle x \rangle}{\ell} =  - \frac{B_p\left[B_p \left(\frac{k_z^2}{2m}-\Delta\right) + B v_F k_y \right]}{B_T^2(\Delta + \Delta')} ,
\end{align}
and extension $\ell_S$. The consistency condition $\langle x \rangle \in [0, \ell]$ leads to
\begin{align}\label{eq:consistencydouble}
	0 > B_p (k_z^2/2m-\Delta) + B v_F k_y > -B_T^2(\Delta + \Delta')/B_p.
\end{align}
The eigenstates are similar to the one defined in (\ref{eq:solrot}) and (\ref{eq:LLzerostate0}). Notice that the $n = 0$ state for vanishing magnetic field on the top surface has the same band dispersion as the Fermi arc found in [\onlinecite{wsmsstheo1}] for a sharp edge, $E_0 = v_F k_y$. This is important insofar as the model used here is quite different with its continuous variation over the step. The band dispersion found in [\onlinecite{wsmsstheo1}] is actually of opposite sign, this is because of different sign conventions on $v_x$ and their results correspond to ours by switching top and bottom surfaces. Notice that beyond this $n=0$ band,
we also find many other gapped ($n \geq 1$) solutions for finite values of the interface width $\ell$ that correspond to higher Landau bands. These do however play no role here in the formation of the Fermi arc, which arises solely from the $n = 0$ band.
We take into account the consistency condition and obtain the following equation for the Fermi surface of the $n = 0$ state at a given chemical potential $\mu$
\begin{align}\label{eq:arcmerging}
	-1 < \frac{\frac{k_z^2}{2m}-\Delta}{\Delta + \Delta'}  + \frac{B}{B_T} \frac{\mu}{\Delta + \Delta'} = \frac{k_y}{\delta k_y} - \frac{ B_p }{ B_T} \frac{\mu}{v_F \delta k_y} < 0.
\end{align}
We represent the resulting Fermi arc in Fig.~\ref{fig:merging} (b) without and, (c) with magnetic field. The external magnetic field bends the Fermi arc which connects the two bulk Fermi surfaces only if $\sqrt{1+(B_p/B)^2} \Delta' > {\rm sign}(B) \mu > \sqrt{1+(B_p/B)^2} \Delta$ and we discuss this in Sec.~\ref{sec:discussion}.

At zero chemical potential, the effect of the magnetic field $B$ is represented by the continuous lines 
in Fig.~\ref{fig:Barcs} for different values of $B$. We observe that the Fermi arc has the equation of a parabola $2 p k_y = k_z^2 - (\Delta k/2)^2$ of parameter $p = v_F m \ell_B^2/\ell_{S}^2$ determined by the magnetic field. This bending can be present without an external magnetic field if we had started with $\delta h_y \neq 0$ which corresponds to a displacement of the bands along $k_y$ from one material to the other. A non-zero value of $\delta h_y$ corresponds to a rotation of the line connecting the Weyl nodes while $\delta h_z$ is responsible for its shortening as described above. Since the coupling of this ``rotation" $\delta h_y$ is precisely that of the applied magnetic field, we can use the expressions derived above simply by replacing $\ell_B=\sqrt{v_x\ell/\delta h_y}$, such that the rotation leads to the same Fermi-arc bending as the magnetic field in Fig.~\ref{fig:Barcs}.

\begin{figure}
    \centering
    \includegraphics[width=\columnwidth]{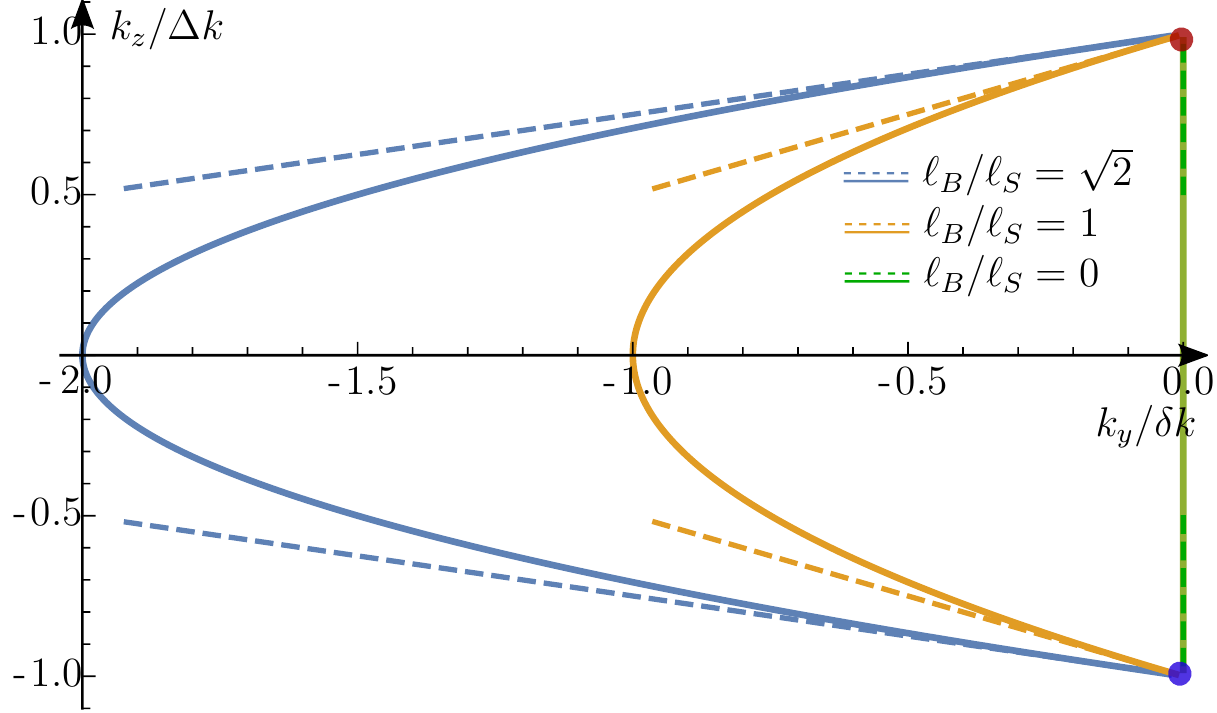}
    \caption{Fermi arc structure of an undoped Weyl semimetal, \emph{i.e.} $\mu = 0$, for different values of the in-plane magnetic field $\mathbf{B} = B\mathbf{e}_z$. Dashed lines are the Fermi arc for independent Weyl cones ($\kappa = 0$) in Eq.~(\ref{eq:Hs0}), the \emph{portions of Fermi arc}, while the continuous lines are for coupled Weyl cones ($\kappa \neq 0$), \emph{full Fermi arcs}.}
    \label{fig:Barcs}
\end{figure}

Notice finally that the deformation of the Fermi arc is also accompanied by a change in the group velocity that should impact transport properties. Indeed, from Eqs. (\ref{eq:LLzerostate0}) and (\ref{eq:LLzerostate0bis}) one finds for the dispersion of the $n=0$ surface band 
\begin{equation}
E_0=v_Fk_y \frac{B_p}{B_T} - \left(\frac{k_z^2}{2m} -\Delta\right)\frac{B}{B_T},
\end{equation}
which yields the group velocity
\begin{align}
{\bf v}_g({\bf B}) = \left(0,v_F\frac{B_p}{B_T},-\frac{k_z}{m}\frac{B}{B_T}\right),
\end{align}
whose orientation depends on the applied magnetic field. In particular, the component along the magnetic field is increased and this may influence magnetotransport as discussed in Ref. [\onlinecite{goswami}] for various metals. The reason is that from Einstein relation [\onlinecite{einstein}] for the conductivity
\begin{align}\label{eq:einstein}
	\sigma =\frac{e^2\tau v_{g,\mathcal{E}}^2}{2}g(\mu),
\end{align} 
with $\tau$ the electrons scattering time, $g(\mu)$ the density of states and ${\bf v}_{g,\mathcal{E}} = {\bf v}_g\cdot\mathcal{E}/\mathcal{E}$ the group velocity along the electric field $\mathcal{E}$, the magnetotransport should depend on the magnetic field. The dependence of magnetotransport on the density of states $g(\omega)$ is discussed in the following section, where we derive the Fermi arc of a single Weyl node which, being isotropic, allows us to study the effect of the magnetic field in any direction.
\subsection{Fraction of a Fermi arc}
\label{sec:fractiont0}

In the previous section we have derived the equation of the Fermi arc  (\ref{eq:arcmerging}) in a magnetic field. In the simplified situation where $B = 0$, one finds $k_y = \mu/v_F$ and $-2m \Delta' < k_z^2 < (\Delta k/2)^2$ ($\Delta' > 0$) which leads to the Fermi arc equation with $k_z \in [-\Delta k/2, \Delta k/2]$. However, one could have considered $\Delta' < 0$ and then the Fermi surface would be cut into two parts $k_z \in [-\Delta k/2, -\sqrt{2m|\Delta'|}]\cup[\sqrt{2m|\Delta'|}, \Delta k/2]$ that we call the fraction of a Fermi arc. This situation describes the interface between two WSM with Weyl nodes located at different momenta in reciprocal space and was observed numerically in Ref. [\onlinecite{wsmssmag1}]. This observation indicates that in the absence of a magnetic field, the Fermi arc only depends on the existence of the insulating gap $\Delta'$ and not on its size. In the case of non-zero in-plane magnetic field, this is still valid at $\mu = 0$. 

In order to obtain additional insight into the Fermi-arc structure and its $B$-field dependence, we now turn to the simplified case of a single Weyl node that shifts in reciprocal space when trespassing the interface. The cones on the left and right sides are described by the following Hamiltonians $\hat{H}_{1L} = \sum_{i \in \{x,y,z\}} v_i k_i \hat{\sigma}_i$ and $\hat{H}_{1R} = \sum_{i \in \{x,y,z\}} v_i (k_i - \delta k_i) \hat{\sigma}_i$. In the notations introduced in section \ref{ssec:interface}, these models are described by the following components
\begin{align}
		\begin{array}{l}
			{\bf h}_{1L}({\bf k}) = v_F (k_x, k_y, k_z),\\
			\\
			{\bf h}_{1R}({\bf k}) = v_F (k_x + \delta k_x, k_y + \delta k_y, k_z + \delta k_z),
		\end{array}
\end{align}
and, for $x \in [0,\ell]$, the interface Hamiltonian is
\begin{align}\label{eq:Hs0}
	\hat{H}_{1s} &= v_F \sum_{i=x,y,z} \left( k_i - \delta k_i x/l\right) \hat{\sigma}_i.
\end{align}
The spatial dependence of this Hamiltonian is similar to what is obtained in a pseudo-magnetic field, $\mathbf{B}_{p} = \nabla \times \mathbf{A}_{p}$, since then the spatial dependence of the Weyl Hamiltonian is $\hat{H}_{\rm el} = \sum_i e v_i A_{p,i}\hat{\sigma}_i$ after a Peierls substitution. In the present situation the pseudo-magnetic field is
\begin{align}\label{eq:Bfield}
	\mathbf{B}_{p} = \delta \mathbf{k} \times \mathbf{e}_x/(e \ell),
\end{align}
where $\mathbf{e}_x$ is the unit vector pointing along the $x-$axis. However, the introduced pseudo-potential vector ${\bf A}_{p}$ is \emph{not} gauge-invariant and $\nabla\cdot{\bf A}_{p} = \delta k_x$, and it is at the origin of the additional $x$-dependent phase $\exp(-ix^2/2\ell_x^2)=\exp(-ix^2\delta k_x/2\ell)$ of the surface states already encountered in Eq. (\ref{eq:wavefon}).

We perform again a rotation of angle $\theta$ of $\hat{H}_{1s}$ along the $x-$axis and obtain a Hamiltonian similar to Eq.~(\ref{eq:Htransfofinale})
\begin{align}\label{eq:rotHs0}
	\hat{H}_{1s}^{(\theta)} &= e^{i\frac{\theta}{2} \hat{\sigma}_x}\hat{H}_{1s}e^{-i\frac{\theta}{2} \hat{\sigma}_x}\\
	&= \left(
	\begin{array}{cc}
		M(k_{\parallel}) & \sqrt{2 v_x v_{p}}~ \hat{a}/\ell_S\\
		\sqrt{2e v_x v_{p}}~\hat{a}^{\dagger}/\ell_S & -M(k_{\parallel})
	\end{array}
	\right),
\end{align}
for $\tan(\theta) = v_z \delta k_z/(v_y \delta k_y)$. We introduce the in-plane velocity $v_p$ perpendicular to ${\bf B}_p$
\begin{align} 
	&v_p = \sqrt{\frac{(v_y \delta k_y)^2 + (v_z \delta k_z)^2}{\delta k_y^2 + \delta k_z^2}},
\end{align}
and $\ell_S = \sqrt{\ell/|\delta {\bf k_{\parallel}}|}$ the effective magnetic length that depends on both the width $\ell$ of the interface and the projection of Weyl node spacing $\delta {\bf k}_{\parallel}$. The mass term $M(k_{\parallel})$ is
\begin{align} \label{eq:E0}
	&M(k_{\parallel}) = \frac{v_y v_z}{v_p} \frac{\left( \delta k_y k_z - \delta k_z k_y\right)}{\sqrt{\delta k_y^2 + \delta k_z^2}}.
\end{align}
The ladder operators are associated with the number states $|n\rangle$ of extension $v_x\ell_s/v_p$, mean position $\langle x \rangle$ and spatial frequency $\lambda_x$ defined by
\begin{align}
	&\frac{\langle x \rangle}{\ell} = \frac{v_y^2 \delta k_y k_y + v_z^2 \delta k_z k_z}{v_p^2 (\delta k_y^2 + \delta k_z^2) },\\
	&\lambda_x = \frac{2\pi}{\delta k_x} \frac{\ell}{\langle x \rangle}.
\end{align}
As we argued in Sec.~\ref{ssec:interface}, in the limit of small $\ell$ one can focus only on the $n = 0$ Landau band defined in Eq.~(\ref{eq:LLzerostate0}). The Fermi surface at chemical potential $\mu $ is then described by 
\begin{align}\label{eq:constraintWeyl}
	0 < \frac{k_y}{\delta k_y} - \frac{v_z \delta k_z}{v_y \delta k_y} \frac{\mu}{v_p |\delta {\bf k}_{\parallel}|} = \frac{k_z}{\delta k_z} + \frac{v_y \delta k_y}{v_z \delta k_z} \frac{\mu}{v_p |\delta {\bf k}_{\parallel}|} < 1,
\end{align}
which reproduces the behavior of a Fermi arc for a single Weyl node. We represent this solution for $\mu = 0$ by the (green) dashed lines in Fig.~\ref{fig:Barcs}. For a better comparison with this case, we double the Fermi arc fractions such as to model two independent Weyl points one situated at $\delta {\bf k}=(0,0,\delta k_z)$ and one at $-\delta {\bf k}$, related by mirror symmetry around $k_z=0$. One clearly sees that the model of independent Weyl points reproduces well the Fermi arcs in the vicinity of the two Weyl nodes. Because of the lack of coupling between the Weyl nodes, the Fermi arcs are broken into two parts, one for each Weyl node, in order to respect the constraint (\ref{eq:constraintWeyl}) on the positions of the surface states. 
This situation has not yet been reported experimentally but was noticed numerically in Ref. [\onlinecite{wsmssmag1}]. We expect that such displacement of Weyl nodes can be observed in the bulk material because of a deformation gradient [\onlinecite{wsmssmag1},\onlinecite{wsmdefo1}-\onlinecite{wsmdefo2}], and the appearance of the previously described metallic states can have a strong influence on bulk dynamics.

We finish this section with a short discussion of the density of states in the Fermi arcs. From the dispersion $E_0(k_y,k_z)= -M(k_y,k_z)$ given by Eq.~(\ref{eq:E0}), where we have replaced $\delta k_y\rightarrow \delta k_y - {\rm sign}(B) \ell/\ell_B^2$ to account for the magnetic field one obtains the $B$-field-dependent density of states per unit area
	\begin{align}\label{eq:DOS}
g_B(\mu) = 
\frac{1}{(2\pi)^2 |v_y v_z|} \sqrt{v_y^2\left[\delta k_y - {\rm sign}(B) \frac{\ell}{\ell_B^2}\right]^2 + (v_z \delta k_z)^2},
	\end{align}
which is independent of the chemical potential $\mu$, $g_B(\mu) = g_B$. In contrast, the bulk Weyl fermion density of states varies quadratically with the chemical potential $g_W(\mu) \sim \mu^2$ and in the limit $\mu \rightarrow 0$ it vanishes, while $g_B$ remains finite. Moreover, one notices from Eq.~(\ref{eq:DOS}) that the magnetic field modifies the surface density of states. Indeed, an expansion around the $B=0$ density of states $g_0 = g_{B=0}$ yields 
	\begin{align}\label{eq:DOS2}
		g_B \approx g_0 - \frac{1}{(2\pi)^2} \frac{|v_y|}{|v_z|}\frac{\delta k_y {\rm sign}(B) \ell/\ell_B^2}{\sqrt{(v_y \delta k_y)^2 + (v_z \delta k_z)^2}},
	\end{align}
i.e. a linear dependence on the magnetic field since $\ell_B\propto 1/\sqrt{B}$. More saliently, depending on the sign of the magnetic field and $\delta k_y$, the density of states can be increased as well as lowered. This should influence the magnetoconductivity derived from the Einstein relation (\ref{eq:einstein}), which is asymmetric under inversion $B \rightarrow -B$ if $\delta k_y \neq 0$. This corresponds to opposite changes in the curvature in the Fermi arc.

Moreover, one can manipulate the displacement of the Weyl nodes with an applied in-plane magnetic field $\mathbf{B}$ of arbitrary direction since the model is isotropic, for example ${\bf B} = B {\bf e}_z$. In the Landau gauge, after Peierls substitution, the applied magnetic field can be described by switching $\delta k_y \rightarrow \delta k_y' = \delta k_y - \textrm{sign}(B)\ell/\ell_B^2$ which translates the Weyl nodes along $y$ in reciprocal space proportionally to the magnetic field $B$. This is plotted in
Fig.~\ref{fig:Barcs} for the same values of the magnetic field as for the previous model with two (merging) Weyl points.
Again the linear dispersion of the Fermi-arc fractions is due to the absence of coupling between the Weyl nodes, but one notices that, in the vicinity of these nodes, the slope of the Fermi arcs is well reproduced within our simplified model. More generally, this deformation of the Fermi arc by a magnetic field is along $\delta {\bf k} = e\ell {\bf e}_x \times {\bf B}$ for each node, independent on its chirality.

In this section, the surface of a Weyl semimetal was modeled with the help of linearly varying model parameters over a finite extension $\ell$. When this size is comparable to the magnetic length $\ell_B$ of an in-plane magnetic field the Fermi arc is deformed. This deformation can be described for any direction of the magnetic field if one considers an isotropic model, which necessarily involves only a single Weyl node and for which we have derived the corresponding Fermi arc. We discussed possible consequences of this finite-sized interface on the contribution of surface states to magnetotransport. In the next section we show that this contribution may not be negligible compared to the bulk response because of the screening of the electric field.\\
In section \ref{sec:electricregime} we extend the description of the surface Hamiltonian to tilted Weyl semimetals and study the surface states of type-I and type-II WSM.


\section{Screening of an external electric field in a Weyl semimetal}
\label{sec:apscreening}
In the previous section we discussed eventual consequences of a magnetic field on the transport properties of the $n = 0$ surface states. From scaling arguments, one can argue that the bulk states dominate transport since $\sigma_{\rm bulk}/\sigma_{\rm surface} \sim L/\Delta x \gg 1$ with $\sigma$ the conductivity, $L$ the sample size and $\Delta x$ the surface states extension. In the present section, we show that the screening of both the $n = 0$ surface state and the bulk states occurs on a length $\ell_{scr.} \lesssim \Delta x$ so that $\sigma_{\rm bulk}/\sigma_{\rm surface} \sim 1$ for a WSM as opposed to the above scaling argument.

In semiconductor physics the electric field at the interface is changed due to charge transfer. We use the results described in [\onlinecite{spiral}] to provide an order of magnitude of the typical depth at which the electric field penetrates in a Weyl semimetal. We consider an electric field along the $y-$axis, then according to the Poisson equation, one has
\begin{align}
	U''(x) = -\frac{e}{\varepsilon_0} n(x),
\end{align}
where $U(x)$ is the electric potential, $e$ is the electron charge and $n(x)$ is the density of electrons.

\subsection{Screening by surface states}\label{ap:ssscreening}
We consider that the Weyl semimetal is located at $x < 0$ and in order to describe the Fermi-arc surface-states we use the same model as in section~\ref{ssec:merging}
\begin{align}
	\hat{H}_L = 	\left(\frac{k_z^2}{2m}-\Delta\right) \hat{\sigma}_z + v_F(k_x \hat{\sigma}_x + k_y\hat{\sigma}_y).
\end{align}
In the following we consider the long-range solution of the $n = 0$ Fermi arc, which corresponds to the behavior for $x \gg \ell$. This solution was derived in Ref. [\onlinecite{sstates}] and we derive it for the finite sized interface in section \ref{sec:topostability}, in Eq.(\ref{eq:longrange}). The long-range solution is such that, for $k_z \in \left[-\Delta k/2, \Delta k/2 \right]$, 
\begin{align}
	&\varepsilon(k_y,k_z) = v_F k_y,\\
	&|\Psi_{s}\rangle = \sqrt{\frac{(\Delta k/2)^2-k_z^2}{m v_F}}\left( 
		\begin{array}{c}
			0\\1
		\end{array}			
	\right)
	e^{\frac{k_z^2-(\Delta k/2)^2}{2 m v_F}x}.
\end{align}
The density of electrons is then 
\begin{align}
	&\frac{\partial n_s}{\partial \mu} = g_s(\mu, x) \\
	&= \frac{1}{(2\pi)^2 v_F} \int_{-\frac{\Delta k}{2}}^{\frac{\Delta k}{2}} dk_z~ \frac{(\Delta k/2)^2-k_z^2}{m v_F}e^{\frac{k_z^2-(\Delta k/2)^2}{m v_F}x}
\end{align}
where $g_s(\mu,x)$ is the local density of states along $x$ per unit of surface.
We rescale $x \rightarrow s = \frac{(\Delta k/2)^2}{m v_F}x$, this way the Poisson equation is
\begin{align}\label{eq:Uscr}
	U''(s) = g \left( \int_{-1}^{1} du~ (1-u^2)e^{s(u^2-1)}\right) U(s),
\end{align}
with $g = 2 \alpha m v_F/\pi \Delta k \approx 4.6$, where $\alpha = e^2/4\pi\varepsilon_r\varepsilon_0 v_F$ is the fine structure constant of the material, in terms of the dielectric constant $\varepsilon_r$ of the material. Here we use the orders of magnitude for {\small $\mathrm{Na_3Bi}$} [\onlinecite{data}]: $\Delta k = 2\sqrt{2m\Delta} = 0.1$ \AA, $v_F = 2.5$ eV\AA~ and $1/2m = 10$ eV\AA$^2$. We observe two characteristic scales
\begin{enumerate}[(i)]

	\item the characteristic depth of surface states $\ell_{scr} = m v_F/(\Delta k/2)^2$. Since the gap $\Delta$ between Weyl cones is $\Delta = (\Delta k/2)^2/2m \approx v_F \Delta k/2$ one can approximate $v_F m/ \Delta k \approx 1$ and $\ell_{scr} \approx 2/\Delta k$.

	\item the characteristic damping amplitude $g = 2 \alpha m v_F/\pi \Delta k \approx 2 \alpha/\pi = (2 \alpha_0/\pi) (c/v_F)\varepsilon_r$, in terms of the bare fine structure constant $\alpha_0=1/137$; for $g > 1$ the damping at the surface is strong and for $g < 1$ the damping at the surface is small.
\end{enumerate}
In the case of {\small $\mathrm{Na_3Bi}$}, $g \approx 4.6$ and $\ell_{scr} \approx 5\textrm{nm}$ and thus we observe a strong decrease of the electric field  due to the $n = 0$ Fermi arc on a length scale of the order of $5\,\textrm{nm}$ as can be seen from numerical calculations Fig.~\ref{fig:surfpot}.

Notice that, here, we have considered the long-range behavior where the interface can be considered sharp with $\ell<1/\Delta k\sim \ell_{scr}$. In this case, the electric field is screened over a small region of depth $\ell_{scr}$ in the WSM. For a wide interface, $\ell \gtrsim 1/\Delta k$, the screening takes place in the interface region and the screening length is then naturally given by the width of the Gaussian wave function $\ell_{scr}\sim \ell_S$ (\ref{eq:wavefon}), which, for a wide interface, is smaller than the interface width $\ell$. In this latter situation, the electric field leads to a renormalization of the surface-state band structure as in the presence of a tilt anisotropy. We discuss this point in more detail in section \ref{sec:electricregime}.

\begin{figure}
	\centering
	\includegraphics[width=0.5\textwidth]{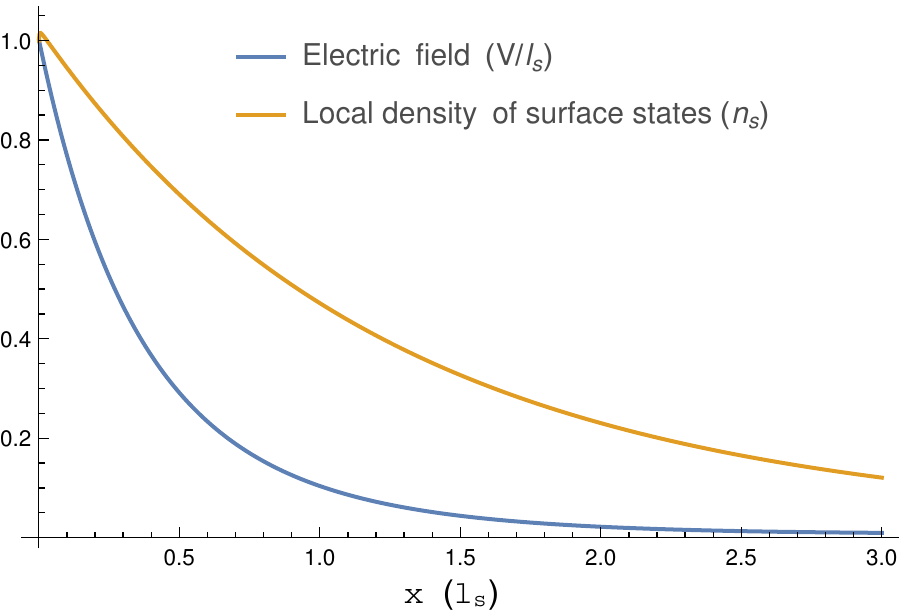}
	\caption{Numerical simulation of the decrease in the electric field at the surface of the Weyl semimetal {\small $\mathrm{Na_3Bi}$} (blue). We also show the local density of states profile (orange) to compare the length scales. One observes the electric field is strongly reduced on the characteristic depth of the Fermi arc surface states.}
	\label{fig:surfpot}
\end{figure}

\subsection{Screening by bulk states}
In the case of $g$ degenerate Weyl cones, with band dispersion $E(\mathbf{k}) = \pm v_F |\mathbf{k}|$, one has the density of states (compressibility) 
\begin{align}
	\frac{\partial n}{\partial \mu} =  g \frac{\mu^2}{\pi^2 v_F^3}.
\end{align}
For an undoped crystal, the electrochemical potential with respect to the Weyl point is $\mu(x) + eU(x) = 0$ and we deduce that
\begin{align}
	U''(x) = -\frac{4g\alpha e^2}{3\pi^2 v_F^2} U^3(x).
\end{align}

We consider the characteristic length to be $\lambda$ and the electric field to be of the order $\mathcal{E} \sim U/\lambda$, then one finds
\begin{align}
	\lambda = \left( \frac{3 \pi^2 v_F^2 }{4 g \alpha e^2 \mathcal{E}^2} \right)^{1/4}
\end{align}

This characteristic length needs to be compared to that, $\ell_{scr}$, intervening in the surface-electron screening. If $\ell_{scr}/\lambda < 1$, the screening is mainly due to the electrons forming the Fermi arcs, while $\ell_{scr}/\lambda >1$ indicates screening primarily due to bulk electrons. One notices that this ratio increases monotonically with the electric field 
\begin{equation}
\frac{\ell_{scr}}{\lambda} =\sqrt{\frac{\mathcal{E}}{\mathcal{E}_0}},
\end{equation}
in terms of the material-dependent characteristic electric field $\mathcal{E}_0 \approx \sqrt{{3}/{g \alpha}}~{\pi \Delta \Delta k}/{16 e}$. 
One therefore notices that at small electric fields $\mathcal{E}<\mathcal{E}_0$ screening is dominated by electrons in the Fermi arcs, while the bulk electrons only contribute to screening at larger values of $\mathcal{E}$. This effect can be understood simply in terms of the density of states -- in spite of their reduced dimension, surface electrons in the Fermi arcs have a non-zero density of states [see Eq.~(\ref{eq:DOS})], while that of the Weyl fermions is strongly reduced ($\sim \mu^2$) in the vicinity of the Weyl nodes. A large electric field is therefore required to induce a sufficient number of bulk carriers such as to have the necessary density of states to provide a significant contribution to screening. 

If we use the same values as for surface states in {\small $\mathrm{Na_3Bi}$}, one obtains a critical field of $\mathcal{E} \approx 10^6 \mathrm{V/m}$, corresponding to a length scale of $\lambda\sim \ell_{scr}\sim 10$ nm.

\section{Surface states of tilted Weyl semimetals}\label{sec:electricregime}

In section \ref{sec:intefmodel}, we derived the Fermi arc structure of type-I WSM with straight cones. 
This situation of straight cones is, however, rarely encountered in materials displaying WSM phases, where the cones are generically tilted. While a moderate tilt barely affects the electronic structure since the isoenergy lines remain closed curves (type-I WSM), the situation is dramatically changed when the cones are overtilted. Parts of the original conical conduction band are then shifted below the Fermi level, while parts of the original valance band float above the Fermi level, leading to open hyperbolic trajectories. First discussed in the framework of 2D organic materials [\onlinecite{goerbig08}] and within a field-theoretical approach for 3D materials [\onlinecite{zubkov14}], these systems are now called type-II WSM [\onlinecite{wsmtype2}] and have been identified experimentally in ARPES [\onlinecite{type2exp1},\onlinecite{type2exp2}]. The tilt and the transition between type-I and type-II WSM have been predicted to have consequences on the Landau bands spectrum [\onlinecite{wsmtilt1}-\onlinecite{wsmtilt3}], namely in magnetooptical spectroscopy. In the present section, we discuss the influence of the tilt on the Fermi arc.

We first discuss the low-energy model we use for type-I and type-II WSM and the corresponding surface Hamiltonian. The surface Hamiltonian appears to be similar to that of electrons in crossed electric and magnetic fields and can therefore be solved with the use of Lorentz boosts. This change in the frame of reference can only be performed if the tilt is smaller than a critical value and above this critical value, the spectrum of surface states collapses.

\subsection{Surface of type-I and type-II WSM}
We model the surface of type-I and type-II WSM with the same procedure as the one we introduced in Sec.~\ref{sec:intefmodel}. The bulk material on the left-side ($x < 0$) of the interface has two tilted Weyl nodes and is in contact with a gapped material ($x > \ell$). We consider that each bulk respects either time-reversal or inversion symmetry such that the two Weyl nodes have opposite tilts [\onlinecite{kawara}]. We describe this situation with the following four-band model
\begin{align}
	\hat{H}_{4t} =&
	 \left( 
	\begin{smallmatrix}
		v_F \mathbf{k} \cdot \left(\hat{\sigma} + \mathbf{t}\mathbbm{1}\right) + \Delta k_0/2 \hat{\sigma}_x & \kappa \mathbbm{1} \\
		\kappa \mathbbm{1} & -v_F \mathbf{k} \cdot \left(\hat{\sigma} + \mathbf{t}\mathbbm{1}\right) + \Delta k_0/2 \hat{\sigma}_x
	\end{smallmatrix}	 \right),
\end{align}
where ${\bf t}$ is the tilt vector. 
This model is similar to Eq.~(\ref{eq:hamqco}) with Weyl nodes located at $k_z = \pm \Delta k/2$ but with opposites tilts. It can describe both types of WSM -- one obtains a 
type-I WSM for $|{\bf t}| < 1$ and a type-II WSM for $|{\bf t}| > 1$ in the case of zero coupling between the Weyl nodes, $\kappa = 0$. 
In Appendix \ref{ap:fourbandreduction} we show that in the limit $|\kappa + v_F \Delta k_0/2| \gg |\kappa - v_F \Delta k_0/2|$, 
this model can be reduced to the following two-band Hamiltonian
\begin{align}\label{eq:twotiltedcones}
	\hat{H}_{2t}(\bf{t},\Delta) &= t_z\left(\frac{k_z^2}{2 m} - \Delta\right) + \frac{2 v_F k_z}{\Delta { k}} (t_x k_x + t_y k_y)\nonumber\\
	&+
	\left(
	\begin{array}{cc}
		\frac{k_z^2}{2 m} - \Delta & v_F (k_x - ik_y)\\
		v_F(k_x + i k_y) & -\left(\frac{k_z^2}{2 m} - \Delta\right)
	\end{array}
	\right).
\end{align}
Strictly speaking, the derivation of this two-band model is valid only for small tilts, $|{\bf t}| \ll 1$. However, we use the two-band model as an effective model here, and lift this constraint in order to also describe type-II WSM for any value of $t_x$, $t_y$ and $t_z$. 
Furthermore, one should keep in mind that this model cannot be used to describe the behavior at larger values of $k_z$, in particular we choose $k_z$ to be on the order of $\Delta k/2$. In the case $\Delta > 0$, one has two Weyl nodes located at $k_z = \pm \Delta k/2 = \pm \sqrt{2 m \Delta}$ with tilts $\pm {\bf t} = \pm (t_x, t_y, t_z)$ (see Fig.~\ref{fig:double_tilted}).  
In the case $\Delta < 0$ one finds a gapped phase that is insulating if $|{\bf t}| < 1$ and if one keeps $k_z$ in the region of size $\Delta k$.

\begin{figure}
	\centering
	\includegraphics[width=\columnwidth]{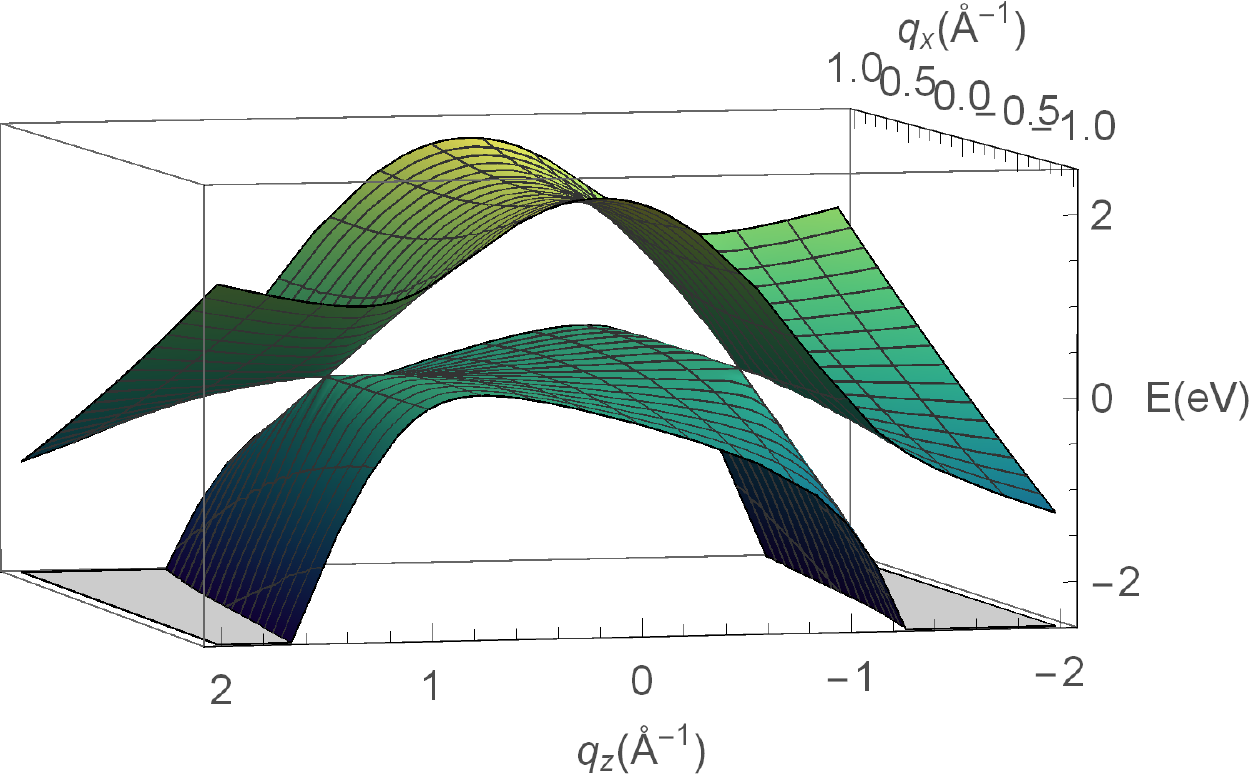}
    \caption{Band dispersion of a type-II WSM obtained from the two band model in Eq.(\ref{eq:twotiltedcones}) at fixed $k_y = 0$.}
    \label{fig:double_tilted}
\end{figure} 

One notices that, as in the absence of a tilt of the Weyl cones, the gap parameter $\Delta$ can be used to trigger the transition from a WSM to an insulating phase and thus to model the interface between the two phases. 
The Hamiltonian for the type-I or type-II WSM on the left side of the interface ($x < 0$) is $\hat{H}_L = \hat{H}({\bf t},\Delta)$ with $\Delta > 0$ and the vacuum ($x > \ell$) is modeled by an insulator of Hamiltonian $\hat{H}_R = \hat{H}({\bf t'}, -\Delta')$ with a gap $\Delta' >0$ and $|{\bf t'}| < 1$.
Within the scheme of linear interpolation between the two bulk Hamiltonians, one obtains the interface Hamiltonian for $x \in [0,\ell]$
\begin{align}
	&\hat{H}_{ts} = \left[ e E_p x + t_z\left(\frac{k_z^2}{2 m} - \Delta\right) + \frac{2v_F k_z}{\Delta k} (t_x k_x + t_y k_y)\right] \mathbbm{1}\nonumber\\
	&+
	\left(
	\begin{array}{cc}
		\frac{k_z^2}{2 m} - \Delta + e v_F B_p x & v_F (k_x - ik_y)\\
		v_F(k_x + i k_y) & -\left(\frac{k_z^2}{2 m}  - \Delta + e v_F B_p x\right)
	\end{array}
	\right),\label{eq:Hts}
\end{align}
with the pseudo-magnetic and pseudo-electric fields
\begin{align}\label{eq:HdoubleBp2}
	&B_p = \frac{\Delta + \Delta'}{e v_F \ell},\\
	&E_p(k_x,k_y,k_z) =\frac{1}{e\ell} \left\{(t_z' - t_z) \frac{k_z^2}{2 m} + t_z \Delta + t_z' \Delta'\right.\\
	 &~~~~~~~~~~~~~~\left.+ \frac{2v_F k_z}{\Delta k}\left[ (t_x' - t_x)k_x + (t_y' - t_y)k_y \right] \right\}.\nonumber
\end{align}
The pseudo-magnetic field $B_p$ was discussed in the framework of Hamiltonian (\ref{eq:HdoubleBp}) but now it competes with the \textit{pseudo-electric field} $E_p$ that has two origins: 
one, $(t_z\Delta+t_z'\Delta')/e\ell$ is due to the change in the gap over the interface and is similar  
to what happens when one introduces the magnetic field with a Peierls substitution for a tilted Weyl node [\onlinecite{wsmtilt3}], and the other one is due to the variation of the tilt in the interface. 
In the case of a surface, where the Hamiltonian $\hat{H}_R$ describes the vacuum, it seems somewhat artificial to have a tilt, and one naturally chooses ${\bf t}'=0$. For formal reasons, we consider nevertheless a surface over which the tilt in the $x$-direction does not change, ${\bf t}' = (t_x, 0, 0)$ -- indeed, the pseudo-electric field depends on momenta and in particular on the out-of-plane momentum $k_x$. This leads to crossed terms in $x$ and $k_x$ that are difficult to study since $[x,k_x] = i$, and in the following we thus consider a pseudo-electric field of the form
\begin{align}\label{eq:Ep}
		E_{p}(k_y,k_z) = - \frac{1}{e\ell}\left[ t_z \left( \frac{k_z^2}{2m} - \Delta\right) + t_y \frac{2 k_z}{\Delta k} v_F k_y \right] .
\end{align}
Notice that, in the presence of a surface gate voltage $V(\ell)$, one can also take into account the voltage drop at the interface by adding a constant $E_0 \approx [V(\ell) - V(0)]/\ell$ to the previous expression of $E_p$. This external electric field $E_0$ is strong for wide interfaces $\ell > 1/\Delta k$, since in this regime voltage drop mostly happens within the interface, as we have argued in Sec. \ref{sec:apscreening}. The pseudo-electric field is oriented along the $x-$axis and competes with the pseudo-magnetic field in the plane of the interface. In this situation, an electron performs a drifted cyclotron motion with drift velocity $v_{D} = E_p/B_p$. One can use a Lorentz boost to eliminate either $E_p$ or $B_p$ depending on the ratio $\beta=v_{D}/v_F$. In the \textit{magnetic regime} where $v_{D} < v_F$, the (pseudo-)magnetic field is dominant, and one can find a frame of reference where the pseudo-electric field vanishes. This regime is characterized by the existence of bound (Landau) states and thus by the presence of surface states. In contrast to this, one obtains an \textit{electric regime} for $v_{D} > v_F$, where only the mapping to a frame of reference without pseudo-magnetic field can be performed. Due to the absence of Landau quantization in this frame of reference, one expects no particular bound states at the interface. If we recall the behavior at ${\bf t} = 0$ in Eq.~(\ref{eq:arcmerging}), we expect from the amplitude of the electric field (\ref{eq:Ep}) that (i), as a consequence of $t_z$, strong effects are observed in the central part of the Fermi arc ($k_z \approx 0$), and (ii), as a consequence of $t_y$, strong effects are observed at $k_z \simeq \pm \Delta k/2$ and for non-zero chemical potential $\mu$. 
In the following we explore this change in frame of reference and show that it also depends on the out-of-plane drift velocity $v_{D,\perp} = 2t_x v_F k_z/\Delta k$ generated by the constant tilt normal to the interface. We would like to stress that the effect of an applied electric field was studied in the context of correlated electrons in [\onlinecite{sselec}] where screening leads to a renormalization that deforms the Fermi arc.

\subsection{Changing the frame of reference}
The surface Hamiltonian (\ref{eq:Hts}) is easier to solve within a convenient frame of reference. We consider no in-plane magnetic field or any shift of the band structure along $k_y$ (\emph{i.e.} $\delta h_y = 0$) and rotate the Hamiltonian by an angle $\pi/2$ along the $x-$axis
\begin{align}
	&\hat{H}_{ts}' = e^{i \pi \hat{\sigma}_x/4}\hat{H}_{ts} e^{-i\pi \hat{\sigma}_x/4}\\
	&~~~~= \left[ e E_p x + t_z\left(\frac{k_z^2}{2 m} - \Delta\right) + \frac{2v_F k_z}{\Delta k} (t_x k_x + t_y k_y) \right]\mathbbm{1}\nonumber \\
	&+
	\left(
	\begin{array}{cc}
		-v_F k_y &  v_F k_x - i e v_F B_p \left(x - \langle x \rangle_0\right)\\
		v_F k_x + i e v_F B_p \left(x - \langle x \rangle_0\right) & v_F k_y
	\end{array}
	\right),\label{eq:rotationtilt}
\end{align}
where $\langle x \rangle_0/\ell = (\Delta - k_z^2/2m)/(\Delta + \Delta')$ is similar to Eq.~(\ref{eq:definitionsFUS}) for $B = 0$. This Hamiltonian has a $k_x-$ and a $x-$dependence on the diagonal that we eliminate with two successive Lorentz boosts (see appendix \ref{ap:hyperbol}). The first Lorentz boost of rapidity $\eta_1$ along the $x$-axis absorbs $2t_x v_F k_z k_x/\Delta k$ and the second Lorentz boost of rapidity $\eta_2$ along the $y$-axis absorbs the pseudo-electric potential $e E_p x$. The combination of these two operations does not correspond to a single Lorentz boost and is accompanied by a rotation, the so-called Thomas-Wigner rotation of relativistic quantum mechanics [\onlinecite{wignerot}]. We derive an expression of the rotation angle in Appendix~{\ref{ap:wignerot}}.
\begin{widetext}
The Schr\"odinger equation $(\hat{H}_{ts}' - E\mathbbm{1})|\Psi\rangle = 0$ becomes
\begin{align}\label{eq:apdefhyp}
	\left( e^{\frac{\eta_2}{2}\hat{\sigma}_y}e^{\frac{\eta_1}{2}\hat{\sigma}_x}\hat{H}_{ts}'e^{\frac{\eta_1}{2}\hat{\sigma}_x}e^{\frac{\eta_2}{2}\hat{\sigma}_y} - E e^{\frac{\eta_2}{2}\hat{\sigma}_y}e^{\eta_1\hat{\sigma}_x}e^{\frac{\eta_2}{2}\hat{\sigma}_y} \right)|\bar{\Psi}\rangle = 0,
\end{align}
where $|\bar{\Psi}\rangle = \mathcal{N} e^{-\frac{\eta_2}{2}\hat{\sigma}_y}e^{-\frac{\eta_1}{2}\hat{\sigma}_x}|\Psi\rangle$ and $\mathcal{N}$ is a normalization constant required since hyperbolic transformations do not preserve the norm of the wave functions. This equation can be written in the form $\hat{H}_{st}^{\eta}|\bar{\Psi}\rangle = E |\bar{\Psi}\rangle$ with
\begin{align}
	\hat{H}_{st}^{\eta} =&  \left[ 2t_y \frac{k_z}{\Delta k} v_F k_y + \left[t_z - T_z({\bf k}_{\parallel})\right]\left(\frac{k_z^2}{2m} - \Delta\right)\right]\mathbbm{1} + \frac{1}{\gamma_3} \left\{ \frac{v_F k_x}{\gamma_1} + \gamma_1 \beta_1 e E_p({\bf k}_{\parallel}) \left[ x - x_1(E, {\bf k}_{\parallel}) \right] \right\}\hat{\sigma}_x\nonumber\\
	& + \frac{1}{\gamma_3}\frac{e v_F B_p }{\gamma_2} \left\{ x - \left[ \langle  x \rangle_0 + \gamma_3^2 T_z^2({\bf k}_{\parallel}) x_1(E, {\bf k}_{\parallel}) \right] \right\}\hat{\sigma}_y - \frac{1}{\gamma_3} v_F k_y \hat{\sigma}_z,
	\label{eq:Htlong}
\end{align}
with $x_0$ is defined in Eq.~(\ref{eq:rotationtilt}), $x_1(E, {\bf k}_{\parallel}) = \left[  E - \frac{2t_y k_z}{\Delta k} v_F k_y - (t_z - T_z({\bf k}_{\parallel}))\left(\frac{k_z^2}{2 m} - \Delta\right)\right]\left/eE_p({\bf k}_{\parallel})\right.$, $ T_z({\bf k}_{\parallel}) = {E_p({\bf k}_{\parallel})}/{(v_F B_p)}$, $\beta_1 = \tanh(\eta_1) = - t_x k_z/\Delta k$, $\beta_2 = \tanh(\eta_2) = -\gamma_1 T_z({\bf k}_{\parallel})$,  and $\gamma_i = 1/\sqrt{1-\beta_i^2}$. 
\end{widetext}
Though this Hamiltonian has a lengthy expression, one notices that the term proportional to $\mathbbm{1}$ does no longer depend on $x$ or $k_x$, i.e. this term does not need to be considered in the diagonalization, while the remaining terms can be diagonalized by the introduction of the usual ladder operators, as we discuss in more detail in the following subsection. 
Notice, at this stage, that the transformation involving the two Lorentz boosts, is only possible if 
\begin{align}\label{eq:conditiontilt}
	\beta_3^2 = \left( 2t_x \frac{k_z}{\Delta k} \right)^2 + T_z \left( {\bf k}_{\parallel} \right)^2 < 1,
\end{align}
to which corresponds $\gamma_3 = \gamma_1 \gamma_2 = 1/\sqrt{1-\beta_3^2} = 1/\lambda_3$. 
This condition states precisely that we are in the magnetic regime, which we mentioned above and which is dominated by the pseudo-magnetic field. 
If for a particular ${\bf k}_{\parallel}$ the condition (\ref{eq:conditiontilt}) is not met, there is no corresponding bound surface state. In the following we solve the surface Hamiltonian for states that fulfill the condition (\ref{eq:conditiontilt}).

\subsection{Surface states of type-I and type-II WSM}
In order to solve Eq.~(\ref{eq:Htlong}), we introduce the ladder operators 
\begin{align}\label{eq:tiltladders1}
	&\hat{b} = \frac{\ell_S}{\sqrt{2} \lambda_3^{3/2} v_F}{\rm Tr}\left[ \hat{H}_{st}^{\eta} \frac{\hat{\sigma}_x-i\hat{\sigma}_y}{2} \right],\\
	\label{eq:tiltladders2}
	&\hat{b}^{\dagger} = \frac{\ell_S}{\sqrt{2} \lambda_3^{3/2} v_F}{\rm Tr}\left[ \hat{H}_{st}^{\eta} \frac{\hat{\sigma}_x+i\hat{\sigma}_y}{2} \right],
\end{align}
which satisfy $[\hat{b},\hat{b}^{\dagger}] = {\rm sign}(v_F \ell)$, and
$\ell_{S} = 1/\sqrt{eB_p}$ is the magnetic length introduced in Eq.~(\ref{eq:HdoubleBp}) for ${\bf t} = 0$ and $B = 0$. We use the ladder operators to rewrite the Hamiltonian (\ref{eq:Htlong}),
\begin{align}
	&\hat{H}_{st}^{\eta} = \left\{ t_y \frac{2k_z}{\Delta k} v_F k_y + \left[t_z - T_z({\bf k}_{\parallel})\right]\left(\frac{k_z^2}{2m} - \Delta\right)\right\}\mathbbm{1} \nonumber\\
	&+ \left(
		\begin{array}{cc}
			- \lambda_3 v_F k_y & \sqrt{2} \lambda_3^{3/2} v_F \hat{b}/\ell_{S}\\
			\sqrt{2} \lambda_3^{3/2} v_F \hat{b}^{\dagger}/\ell_{S} & \lambda_3 v_F k_y
		\end{array}
	\right).
\end{align}
The eigenvalues of this Hamiltonian are again similar to Landau bands. In the case $v_x \ell > 0$, $[\hat{b},\hat{b}^{\dagger}] = 1$ and the eigenstates can be written in the form
\begin{align}
	|\Psi\rangle = \left(
		\begin{array}{c}
			\alpha_n |n-1\rangle\\
			\beta_n |n\rangle
		\end{array}
	\right),
\end{align}
where $|n\rangle$ are the eigenstates of the number operator $\hat{n} = \hat{b}^{\dagger}\hat{b}$ of eigenvalue $n \in \mathbb{N}$. The $|n\rangle$ states explicitly depend on the energy and, in particular, we find that the mean position is
\begin{align}
	 &\langle x \rangle = \langle  x \rangle_0 + \gamma_3^2 T_z^2({\bf k}_{\parallel}) x_1(E, {\bf k}_{\parallel}),
	 \label{eq:meanpost}
\end{align}
which must again be located in the interval $[0,\ell]$. 

Since the average position depends on the energy, the consistency condition (\ref{eq:consistencydouble}) now also depends on the quantum number $n$. The $n \geq 1$ band are still gapped with eigenstates in the original basis $|\Psi_{\sigma,n}\rangle$ and eigenenergies $E_{\sigma,n}$ ($\sigma = \pm 1$) such that
\begin{align}\label{eq:LLntilt}
		\begin{array}{l}
			|\Psi_{\sigma,n}\rangle = 
			\hat{R} \left( 
				\begin{array}{c}
					\left( 1 - \sigma \frac{\lambda_3 v_F k_y}{E_n} \right) | n - 1 \rangle \\
					\sigma \left( 1 + \sigma \frac{\lambda_3 v_F k_y}{E_n} \right)|n\rangle
				\end{array}
			\right),\\~~\\
		E_{\sigma,n}(\mathbf{k}_{\parallel}) = t_y \frac{2k_z}{\Delta k} v_F k_y + \left(t_z - T_z({\bf k}_{\parallel})\right)\left(\frac{k_z^2}{2m} - \Delta\right) + \sigma E_n,
		\end{array}
\end{align}
with $\hat{R} = \mathcal{N} e^{-\eta_2\hat{\sigma}_y/2}e^{-\eta_1\hat{\sigma}_x/2}e^{i \pi \hat{\sigma}_x /4}$ and $E_n = \sqrt{ \lambda_3^2 v_F^2 k_y^2 + 2 \lambda_3^{3} v_F^2 n/\ell_S^2}$.
These solutions are similar to those obtained in Eq.~(\ref{eq:solrot}) but with a renormalization of the $n \geq 1$ gaps by a factor $(1-\beta_3^2)^{3/4}$, due to the abovementioned succession of two hyperbolic transformations. Because of this renormalization, which can be further enhanced by the application of a true electric field perpendicular to the surface, we expect that the $n \geq 1$ surface states can be seen experimentally for some surface orientations, since $\beta_3$ is roughly the tilt component transverse to the pseudo-magnetic field whose orientation depends on the surface cut.

\begin{figure}[t]
    \centering
    \includegraphics[width=\columnwidth]{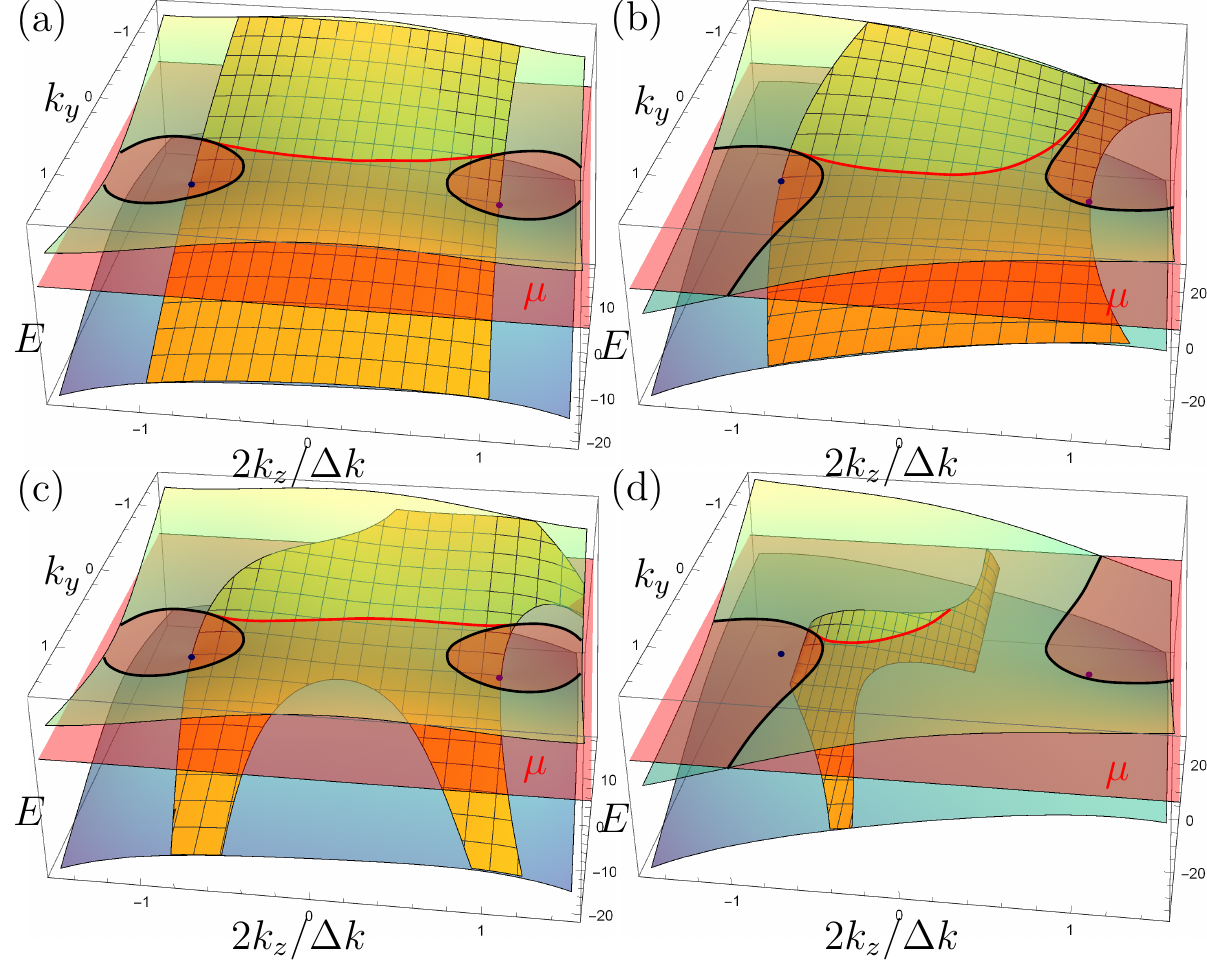}
    \caption{Superposition, similar to Fig.~\ref{fig:merging} (b) and (c), of the bulk WSM band structure at $k_x = 0$ (tranparent green and blue), the band dispersion of $n = 0$ surface states (orange) and the constant energy surface $\mu$ (transparent red). In (a) and (b) the insulating gap is large $\Delta' \gg \Delta$ while in (c) and (d) it is $\Delta' \sim \Delta$. Apart from the insulating gap, (a) and (c) correspond to a type-I WSM with ${\bf t}_1 = (0.1,0.1,-0.4)$ ($|{\bf t}_1| = 0.18 < 1$) while (b) and (d) correspond to the a type-II WSM with ${\bf t}_2 = (0.1,0.75,-0.75)$ ($|{\bf t}_2| = 1.13 > 1$). In the limit of a large gap insulator ($\Delta' \gg \Delta$), the Fermi arc is always connecting the two Fermi pockets. However, for a smaller insulating gap ($\Delta' \sim \Delta$), while the type-I WSM still has the Fermi arc connecting the two pockets, in the type-II WSM one part of the Fermi arc can be dangling between the two nodes.}
    \label{fig:tiltedFermiarc}
\end{figure}

The more interesting $n = 0$ surface state is also present, and it is still observable in the limit $\ell \rightarrow 0$ where the $n \geq 1$ gaps become infinitely large,
\begin{align}\label{eq:LLzerostatetilt}
		\begin{array}{l}
			|\Psi_{0}\rangle = 
			\mathcal{N} e^{-\eta_2\hat{\sigma}_y/2}e^{-\eta_1\hat{\sigma}_x/2}e^{i \pi \hat{\sigma}_x /4} \left( 
				\begin{array}{c}
					0 \\
					|0\rangle
				\end{array}
			\right),\\~~\\
		E_{0}(\mathbf{k}_{\parallel}) = t_y \frac{2k_z}{\Delta k} v_F k_y + \left[t_z - T_z({\bf k}_{\parallel})\right]\left(\frac{k_z^2}{2m} - \Delta\right)  - \lambda_3 v_F k_y
		\end{array}
\end{align}
This solution leads to the $n = 0$ Fermi arc that we represent in Fig.~\ref{fig:tiltedFermiarc} for two different tilts. While the panels (a) and (c) on the left hand side represent the situation of a type-I WSM with a moderate tilt, the panels (b) and (d) on the right hand side correspond to a type-II WSM. In order to obtain further insight into the Fermi-arc structure, we show the case of a surface (with $\Delta'\gg \Delta$) in the upper panels (a) and (b), in comparison with an interface between a WSM and a small-gap insulator ($\Delta'\ll \Delta$) in the lower panels (c) and (d). 

We stress that for tilted WSM the existence of bound states is subject to two conditions (i) the condition on localization  $\langle x \rangle \in [0,\ell]$ we discussed in Eq.~(\ref{eq:consistencygeneral}) and, (ii) the maximal-boost condition for being in the magnetic regime $|\beta_3| < 1$ we derived in Eq.~(\ref{eq:conditiontilt}). In the limit $\Delta' \gg \Delta$, we recover the localization condition (\ref{eq:arcmerging}) as for the case of no tilt because
 $t_x < 1$ and $T_z({\bf k}_{\parallel}) \ll 1$, and one obtains a full Fermi arc
unless a very strong magnetic field (or a component $\delta h_y$) is applied to the system, $\ell_B\lesssim\ell_S$, and for strong doping, $\mu\sim \Delta$ as in Fig.~(\ref{fig:merging}). In this rather theoretical case, the Fermi arc can be covered by the bulk Fermi sea (not shown here).
In the case of an interface between a WSM and a small-gap insulator, i.e. $\Delta' \ll \Delta$, the two cones are well separated from each other, and the maximal-boost condition (\ref{eq:conditiontilt}) may then be approximated by 
\begin{align}\label{eq:limittiltgap}
	\left( 2t_x \frac{k_z}{\Delta k} \right)^2 + \left( \frac{ t_z(k_z^2 - (\Delta k/2)^2) + t_y (2k_z/\Delta k) 2m v_F k_y }{(\Delta k/2)^2}\right)^2 < 1.
\end{align}
Furthermore, the localization condition for the $n = 0$ Landau band is
\begin{widetext}
\begin{align}\label{eq:conditionwithtilt}
	- 2 m \Delta' < \left( 1 - s \frac{v_F k_y}{\lambda_3(\Delta + \Delta')} t_z \right) k_z^2 + s \frac{v_F k_y}{\lambda_3(\Delta + \Delta')} \left( t_z (\Delta k/2)^2 - t_y \frac{2k_z}{\Delta k} 2 m v_F k_y  \right) < \Delta k^2,
\end{align}
\end{widetext}
with $s = {\rm sign}(v_x \ell)$. In order to understand in more detail these two conditions, consider the simplified case $t_y = 0$. If $t_y = 0$, one obtains from Eq.~(\ref{eq:limittiltgap}) and Eq.~(\ref{eq:conditionwithtilt}), with $E_0({\bf k}_{\parallel}) = \mu$, the inequalities
\begin{align}
	&t_x^2 q_z^2 + t_z^2 \left[ q_z^2 - 1 \right]^2 = \beta_3^2 < 1,\\
	& - \frac{1 - t_x^2 q_z^2 + t_z \left[ t_z \left( q_z^2 - 1 \right) - \frac{\mu}{\Delta}\right]}{ 1 - t_x^2 q_z^2 - t_z\left[ t_z (2 q_z^2-1)\left( q_z^2 - 1 \right) - \frac{\mu}{\Delta}\right]} < q_z^2 < 1,
\end{align}
respectively, where $q_z = 2k_z/\Delta k$. These two conditions show that the effect of $t_z$ is to cut the Fermi arc, even at $\mu = 0$ for $t_z > 1$ (type-II WSM) and the arc totally disappears once $t_x^2(1-t_x^2/4t_z^2) > 1$. This is shown in Fig.~\ref{fig:tiltedFermiarc} (d), where one notices a shortening of the Fermi arc that does no longer connect the electron pockets. Moreover, for large doping (of the order $\Delta/|{\bf t}|$) the tilt destroys the $n = 0$ Fermi arc. Through these observations we would like to stress that, contrary to type-I WSM, for type-II WSM one can encounter a situation where the Fermi arc does not connect the bulk Fermi pockets for finite values of the insulating gap $\Delta'$.
One also notices that, for a tilt in the $x$-direction with $t_y=0$, the disappearance of the Fermi arc depends on the surface orientation. Had we considered a surface with a normal vector along $y$, we would have to interchange $t_x\leftrightarrow t_y$, and the corresponding condition is never satisfied. Therefore, the Fermi arc could be visible on one surface but not on another one with a different orientation.

Notice that the $n=0$ Fermi arc depends on the surface orientation defined by ${\rm sign}(v_x \ell)$. For arbitrary ${\rm sign}(v_x \ell)$, the isospin of the $n = 0$ state points along the direction
\begin{align}
	\label{eq:polaf}
	{\bf v}_t({\bf k}_{\parallel}) &= - \gamma_1 \beta_1 {\bf e}_x - \gamma_2 [\gamma_1 \beta_2 + {\rm sign}(v_x\ell)] {\bf e}_y
\end{align}
and is of eigenenergy $E_{0}({\bf k}_{\parallel}) = 2t_y (k_z/\Delta k) v_F k_y + [t_z - T_z({\bf k}_{\parallel})](k_z^2/2m - \Delta) + {\rm sign}(v_x\ell)\lambda_3 v_F k_y$. At a fixed ${\bf k}_{\parallel}$ one observes that opposite surface cuts have different orientation of the iso-spin ${\bf v}_t$. However for time-reversal invariant materials there is no net magnetization since opposite points $\pm{\bf k}_{\parallel}$ in the reciprocal space have opposite tilts [\onlinecite{kawara}]. We finally emphasize that in the limit $\Delta' \ll \Delta$, no features are expected from Landau bands with $n \neq 0$ if their energy spacing $\sqrt{\Delta/v_F \ell}$ is larger than $\Delta'$. If we consider for instance a characteristic gap $\Delta' = 100$ meV and a Fermi velocity $v_F = 2.5$ eV\AA~(as for Na$_3$Bi), one finds 
that for $\ell < v_F/\Delta' \sim 2.5$ nm, the $n\neq 0$ Landau bands overlap in energy with the bulk states of the insulator. 

\section{Topological stability}\label{sec:topostability}
The Fermi arc stability is sometimes related to the Chern number of a 2D topological insulator with the $k_z$-dependence playing the role of a mass term [\onlinecite{dlc}]. In this section we use the analogy of surface states with Landau bands to discuss their stability. This can be achieved with the help of an adapted Aharonov-Casher argument [\onlinecite{acasher}]. The original argument demonstrates that the $n=0$ Landau band of the Dirac equation is independent of fluctuations in the magnetic field. This argument is related to existence of a chirality operator and was also demonstrated for tilted 2D Dirac systems in Ref. [\onlinecite{haoki1}]. In the following we discuss the notion of chirality for 3D tilted band structures. This chirality is used to demonstrate the topological stability of the Fermi arc for both type-I and type-II WSM.

\subsection{Generalized chirality}
The concept of chirality as a matrix that anti-commutes with the Hamiltonian is of fundamental importance when discussing localization in a space-dependent potential [\onlinecite{haoki1}]. On general grounds, one can consider a Hamiltonian of the form
\begin{align}\label{natilt}
	\hat{H}_{h_0,\mathbf{h}} = h_0 \mathbbm{1} + \mathbf{h}\cdot \hat{\mathbf{J}}~~,
\end{align}
where $\mathbf{\hat{J}}$ are Hermitian matrices and one can write $h_0 = \mathbf{t}\cdot {\bf h}$ for a tilted dispersion relation with the tilt parameters ${\bf t}$. If one considers a spatial variation of $(h_0,\mathbf{h})$, for example $(h_0(x),\mathbf{h}(x)) = (h_0+ \delta h_0(x),\mathbf{h}+ \delta \mathbf{h}(x))$ for a planar interface, one can separate the Schr\"odinger equation $(\hat{H}_{h_0,\mathbf{h}}-E\mathbbm{1})|\Psi\rangle = 0$ into a part that explicitly depends on position and one part that does not. In the previous example, one can separate an operator depending on $k_x$ and $x$ from one that depends on $E$, $k_y$ and $k_z$,
\begin{align}
	\hat{H}_x|\Psi\rangle = \hat{H}_{E,yz}|\Psi\rangle
	\label{nschro}
\end{align}
where $\hat{H}_{h_0,\mathbf{h}}-E\mathbbm{1} = \hat{H}_x - \hat{H}_{E,yz}$.

Let us consider for a moment that we already know a Hermitian chirality operator $\hat{\kappa}$ such that $\hat{\kappa}^2 = 1$ and that it anti-commutes with $\hat{H}_{x}$,
\begin{align}
	\left\{ \hat{\kappa} , \hat{H}_{x}\right\} = 0.
	\label{antico}
\end{align}
We explicitly construct an example of $\hat{\kappa}$ at the end of this subsection. We project Eq. (\ref{nschro}) on the basis of eigenstates $|\pm\rangle$ of $\hat{\kappa}$ and write as in [\onlinecite{haoki1},\onlinecite{alternant}]
\begin{align}
\left[
	\begin{array}{cc}
		\langle + | \hat{H}_{x} | +\rangle & \langle + | \hat{H}_{x} | -\rangle\\
		\langle - | \hat{H}_{x} | +\rangle & \langle - | \hat{H}_{x} | -\rangle
	\end{array}
	\right]&
	\left[
		\begin{array}{c}
			\Psi_+\\
			\Psi_-
		\end{array}
	\right] \nonumber\\= 
	\left[
	\begin{array}{cc}
		\langle + | \hat{H}_{E,yz} | +\rangle & \langle + | \hat{H}_{E,yz} | -\rangle\\
		\langle - | \hat{H}_{E,yz} | +\rangle & \langle - | \hat{H}_{E,yz} | -\rangle
	\end{array}
	\right]&
	\left[
		\begin{array}{c}
			\Psi_+\\
			\Psi_-
		\end{array}
	\right].
\end{align}
Since $\hat{H}_x$ is of chiral form in the $|\pm\rangle$ basis,
$\langle + | \hat{H}_{x} | +\rangle = 0 = \langle - | \hat{H}_{x} | -\rangle$, one obtains the generic form
\begin{align}
	\left[
	\begin{array}{cc}
		0 & \hat{\alpha}_-\\
		 \hat{\alpha}_+ & 0
	\end{array}
	\right]&
	\left[
		\begin{array}{c}
			\Psi_+\\
			\Psi_-
		\end{array}
	\right] = 
	\left[
	\begin{array}{cc}
		\langle + | \hat{H}_{E,yz} | +\rangle & \langle + | \hat{H}_{E,yz} | -\rangle\\
		\langle - | \hat{H}_{E,yz} | +\rangle & \langle - | \hat{H}_{E,yz} | -\rangle
	\end{array}
	\right]
	\left[
		\begin{array}{c}
			\Psi_+\\
			\Psi_-
		\end{array}
	\right],
\end{align}
where $\hat{\alpha}_-(x,k_x)$, $\hat{\alpha}_+(x,k_x)$ play the role of ladder operators in our calculations. One can search for solutions of the form $\Psi_1 = (0, \Psi_-)$ and $\Psi_2 = (\Psi_+, 0)$ and then find the corresponding zero-state equations [\onlinecite{haoki1}]
\begin{align}\label{ap:zeromodes}
	\Psi_1 : &\left\{
	\begin{array}{l}
	\hat{\alpha}_- \Psi_- = \langle + | \hat{H}_{E,yz} | -\rangle \Psi_-\\
	\langle - | \hat{H}_{E,yz} | -\rangle = 0
	\end{array}
	\right.,\\
	\Psi_2 : &\left\{
	\begin{array}{l}
	\hat{\alpha}_+ \Psi_+ = \langle - | \hat{H}_{E,yz} | +\rangle \Psi_-\\
	\langle + | \hat{H}_{E,yz} | +\rangle = 0
	\end{array}
	\right. .
\end{align}
These states have topological properties that we discuss in the next section, using an argument similar to the Aharonov-Casher argument as in Refs. [\onlinecite{acasher},\onlinecite{haoki1},\onlinecite{haoki2}].

In the case of a tilted Hamiltonian relation Eq.~(\ref{antico}) cannot be satisfied because then $\delta h_0 \neq 0$ or $h_0$ have an explicit $k_x$-dependence, and some terms proportional to the identity, $\mathbbm{1}$, appear in $\hat{H}_x$. However, as in Refs. [\onlinecite{kawara},\onlinecite{haoki1}, \onlinecite{haoki2}] a generalized chirality operator can be defined under certain circumstances with help of hyperbolic transformations. The matrix $\hat{\kappa}$ does not need to be Hermitian and instead of the anti-commutation relation Eq.~(\ref{antico}) one needs [\onlinecite{alternant},\onlinecite{haoki1}]
\begin{align}
	\hat{\kappa}^{\dagger} \hat{H}_x \hat{\kappa} = - \hat{H}_x,
\end{align}
where $\hat{\kappa}$ should be diagonalizable and such that $\hat{\kappa}^2 = 1$ (\emph{i.e.} $\hat{\kappa}$ is an involutory matrix).

The use of hyperbolic transformations introduced in Appendix \ref{htr} allows us to find a generalized chirality, $\hat{\kappa}_{\theta}$, for Eq.~(\ref{natilt}). We consider that $\hat{H}$ for a given $h_0$ and $\delta h_0(x)$ can be related to a $\hat{H}' = e^{\theta \hat{\Gamma}}\hat{H}e^{\theta \hat{\Gamma}}$ with a known chirality $\hat{\kappa}$ as in Eq.~(\ref{antico}) using a hyperbolic transformation. Typically $\hat{H}'$ is such that $h_0' = h_0(k_x = 0)$ and $\delta h_0'(x) = 0$. The transformation is of the generic form
\begin{align}
	|\Psi ' \rangle = N e^{-\theta \hat{\Gamma}} |\Psi \rangle,
\end{align}
where $N$ is a normalization constant. Then Eq.~(\ref{nschro}) becomes
\begin{align}
	e^{\theta \hat{\Gamma}}\hat{H}_{x}e^{\theta \hat{\Gamma}} |\Psi\rangle = e^{\theta \hat{\Gamma}}\hat{H}_{E,yz}e^{\theta \hat{\Gamma}} |\Psi\rangle
\end{align}
and since the chirality operator associated to $\hat{H}_x' = e^{\theta \hat{\Gamma}}\hat{H}_{x}e^{\theta \hat{\Gamma}}$ is $\hat{\kappa}$, one finds
\begin{align}
	&\left\{ \hat{\kappa} , e^{\theta \hat{\Gamma}}\hat{H}_{x}e^{\theta \hat{\Gamma}}\right\} = 0\\
	\implies &\kappa_{\theta}^{\dagger} \hat{H}_{x} \kappa_{\theta} = - \hat{H}_{x}
\end{align}
with ${\kappa}_{\theta} = e^{\theta \hat{\Gamma}}\hat{\kappa}e^{-\theta \hat{\Gamma}}$ the chirality operator for the Hamiltonian $\hat{H}_{x}$. 

We can illustrate these general considerations with the model (\ref{eq:Hsdoub}), which describes the interface between two Weyl nodes and an insulator,
\begin{align}\label{eq:Htiltchiral}
	\hat{H}_2(t_z) &= v_F \left( k_x \hat{\sigma}_x + k_y \hat{\sigma}_y \right)\\
	 &~~~~~~~~~~~~~~~~+ \left( \frac{k_z^2}{2m} - \Delta \right)(t_z\mathbbm{1} + \hat{\sigma}_z)\nonumber,
\end{align}
with $\delta \mathbf{h}(x) = (0,0,2 \Delta x/\ell)$. The chiral part of the Hamiltonian depending on $x$ and $k_x$ then reads
\begin{align}\label{example:Hx}
	\hat{H}_{2x}(t) = v_F k_x \hat{\sigma}_x + \Delta(2x/\ell -1)(t_z \mathbbm{1} + \hat{\sigma}_z).
\end{align}
In the case $t_z = 0$, the corresponding chirality is $\hat{\kappa} = \hat{\sigma}_y$ and Eq.~(\ref{example:Hx}) is linked to this operator simply by the hyperbolic transformation $\exp(\eta\hat{\sigma}_x/2)$
\begin{align}
	\hat{H}_{2x}(t_z = 0) = e^{\frac{\eta}{2} \hat{\sigma}_x}\hat{H}_{2x}(t) e^{\frac{\eta}{2} \hat{\sigma}_x},
\end{align}
with $\tanh(\eta) = \beta =  - t_z \in [-1,1]$. The chirality operator corresponding to $\hat{H}_{2x}(t_z)$ is thus
\begin{align}
	\hat{\kappa}_t &= e^{\frac{\eta}{2} \hat{\sigma}_x}\hat{\sigma}_y e^{-\frac{\eta}{2} \hat{\sigma}_x}\\
	&= \gamma \left( \hat{\sigma}_y - i \beta \hat{\sigma}_x \right)
\end{align}
with $\gamma = (1-\beta^2)^{-1/2}$. The use of hyperbolic rotations allows us to express the chirality matrix for a broader class of systems as in Ref. [\onlinecite{kawara}]. It also shows that for an overtilted band dispersion in the $z$-direction ($|t_z| > 1$), it is not possible to define a chirality operator. Notice, however, that the argument still holds in the case of a type-II WSM with a prominent tilt in a direction perpendicular to the interface and to the direction connecting the Weyl cones. Indeed, if we had replaced in Eq.~(\ref{eq:Htiltchiral}) $\hat{\sigma}_y$ by $t_y1\!\! 1+ \hat{\sigma}_y$, one could still find a Lorentz boost to a frame of reference where the $x$- and $k_x$-dependent part of the Hamiltonian is cast into a chiral form, and one would then obtain stable Fermi arcs as long as $|t_z|<1$. This also shows, as we have already mentioned above, that the presence or absence of a Fermi arc in a type-II WSM depends on the surface orientation.

\subsection{The Aharonov-Casher argument for surface states}

In the previous sections, we have introduced toy models of surface Hamiltonians that linearly interpolate between the two bulk Hamiltonians $\hat{H}_L$, at $x < 0$, and $\hat{H}_R$, at $x > \ell$, for $x \in [0,\ell]$. This corresponds to a uniform in-plane magnetic field and one can wonder if the surface states are robust to inhomogeneities in the surface potential as illustrated schematically in Fig.~\ref{fig:randominterface}.
 
\begin{figure}[thb]
    \centering
    \includegraphics[width=\columnwidth]{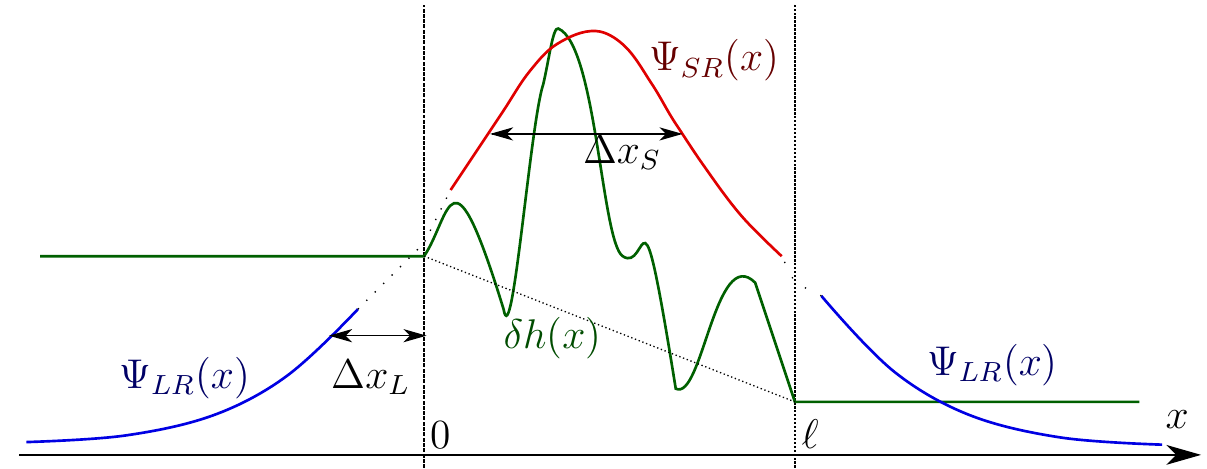}
    \caption{The interface potential can be inhomogeneous over the size $\ell$ as pictured on this figure. In the short-range limit, the wavefunction $\Psi_{SR}(x)$ (in red) strongly depends on the interface potential (in green) and in the case of a linear profile it is a Gaussian of characteristic length $\Delta x_S = 1/\sqrt{eB_p'} \sim \sqrt{\ell}$. The long-range solution $\Psi_{LR}(x)$ (in blue) consists of decaying exponentials with a characteristic length $\Delta x_{L} = 1/(eB_p'\ell)\sim \ell^0=1$. }
    \label{fig:randominterface}
\end{figure}

In the following, we consider an interface Hamiltonian of the form (\ref{natilt}) that interpolates between $\hat{H}_L$ and $\hat{H}_R$ with a fixed orientation $\left[ \Delta h_0, \Delta \mathbf{h}\right] = \left[ \frac{1}{2}\mathrm{Tr}\left((\hat{H}_R - \hat{H}_L)\right),\frac{1}{2}\mathrm{Tr}\left((\hat{H}_R - \hat{H}_L){\hat{\boldsymbol \sigma}}\right) \right]$ but with an arbitrary profile
\begin{align}
	\hat{H}_s = \hat{H}_L + \delta h_0(x) \mathbbm{1} + \delta {\bf h}(x) \cdot \hat{\boldsymbol{\sigma}},
\end{align}
with $(\delta h_0(x), \delta {\bf h}(x)) =  f(x) (\Delta h_0, \Delta {\bf h})/\ell$ and where the function $f(x)$ is chosen such that $f(x=0)=0$ and $f(x=\ell)=\ell$. One can associate pseudo-electric and pseudo-magnetic fields ${\bf e}_p$ and $\mathbf{b}_p$ such that
\begin{align}
	&\mathbf{e}_p = \partial_x \delta h_0(x) {\bf e}_x,\\
	&\mathbf{b}_p = {\boldsymbol\nabla} \times \delta \mathbf{h}(x) = \mathbf{e}_x \times \partial_x \delta \mathbf{h}(x),
\end{align}
for which the associated one-dimensional flux is
\begin{align}
	\int_{-\infty}^{\infty} \mathbf{b}_p \times \mathbf{e}_x dx = \int_{-\infty}^{\infty} \partial_x \delta \mathbf{h}_{\parallel} dx = \Delta \mathbf{h}_{\parallel}.
\end{align}

In order to prove that the interface has a unique zero-mode eigenstate (\ref{ap:zeromodes}) with spin orientation dependent on $\Delta {\bf h}_{\parallel}$, we consider that the only $k_x$-dependence is in $h_0 = t_x h_x$, $h_x = v_x k_x$. Then we perform (i) a rotation $e^{i\theta \hat{\sigma}_x/2}$ to transform $(\delta h_x, \delta h_y, \delta h_z) \rightarrow (\delta h_x, \delta \tilde{h}_y, 0)$ as in Eq.~(\ref{eq:rotationtilt}) with $\tan(\theta) = \delta h_z/\delta h_y$ and, (ii) two Lorentz boosts $e^{\eta_1 \hat{\sigma}_x/2}$, $e^{\eta_2 \hat{\sigma}_y/2}$ to absorb the $k_x$ and $x$ dependence in $h_0 + \delta h_0(x)$ as in Eq.(\ref{eq:apdefhyp}) with $\tanh(\eta_1) = \beta_1$ and $\tanh(\eta_2) = \beta_2$. The transformations are still position-independent because each term scales with the same $f(x)$. The Hamiltonian after the transformation is of the form
\begin{align}
	\hat{H}_s' = h_0' \mathbbm{1} + {\bf h}'(x) \cdot \hat{\boldsymbol{\sigma}},
\end{align}
with $h_x' = v_F' k_x + eE_p'[f(x) - x_0]$, $h_y' = e v_F' B_p'[f(x) - x_1]$ and $h_z'(x) = h_z'$ as in Eq.(\ref{eq:Htlong}). Here, the primes at  the pseudo-electric $E_p'$ and the pseudo-magnetic fields $B_p'$ indicate their respective values in the Lorentz-transformed frame of reference and in particular 
\begin{align}
	e v_F' B_p' = \left.{\sqrt{\Delta h_y^2 + \Delta h_z^2}}\right/{\gamma_3 \gamma_2}.
\end{align} 
Furthermore, the parameters $x_0$ and $x_1$ correspond to the shifts introduced in Eq.~(\ref{eq:Htlong}). We omit the precise form of $x_0$ which is not relevent here, and find for $x_1$
\begin{align}
	x_1/\ell = &- \frac{\Delta {\bf h}_{\parallel} \cdot {\bf h}_{L}}{\Delta {\bf h}_{\parallel}^2}\\
	&+ \gamma_3^2 \left( \frac{\Delta h_0}{\Delta h_{\parallel}} \right)^2 \left[ E - h_{0}(k_x = 0) + \frac{\Delta h_0}{\Delta h_{\parallel}} h_3 \right],\nonumber
\end{align} 
which is similar to the expression for the mean position in Eq.~(\ref{eq:meanpost}).
The $x-$position dependent part is then
\begin{align}
	\hat{H}_{sx}' = \left\{v_F' k_x + V[f(x) - x_0]\right\} \hat{\sigma}_x& \nonumber\\
	+ e v_F' B_p'[f(x) - x_1]\hat{\sigma}_y,&
\end{align}
with an associated chirality operator $\kappa = \hat{\sigma}_z$, zero modes $\Psi_{\sigma}$, $\sigma = \pm$ such that $\Psi_{\sigma} = \psi_{\sigma} | \sigma \rangle_z$  [as in Eq.~(\ref{ap:zeromodes})] and
\begin{align}\label{eq:zero}
		\begin{array}{l}
			\left\{v_F' \partial_x + i eE_p'[f(x) - x_0] + \sigma e v_F' B_p'[f(x) - x_1] \right\}\psi_{\sigma} = 0,\\
			E_{\sigma} = h_0' + \sigma h_z'.
		\end{array}
\end{align}
Its solutions are of the form $\psi_{\sigma} = e^{\chi_{\sigma}}$ with
\begin{align}
	\chi_{\sigma} = \chi_{\sigma}^{(0)} - e \sigma B_p'\int_{0}^{x} dx' \left[f(x) - x_1\right]\nonumber &\\
	 - i \frac{eE_p'}{v_F} \int_{0}^{x} dx' \left[f(x) - x_0\right],&
\end{align}
where one can show with the help of an integration by parts that
\begin{align}
	\int_{0}^{x} dx' g(x') = xg(x=0) + \int_{0}^{x}dx' (x - x') \partial_{x'}g(x').
\end{align}
We consider that $\delta {\bf h}(x)$ is mostly varying in the interface $x \in [0,\ell]$ and that we are only interested in the long-range behavior of the solution, far away from the interface where $\partial_x f(x) \ll 1$. Then far away from the interface one can write
\begin{align}
	\int_{0}^{x}dx' (x - x') \partial_{x'} f(x') \approx \Theta(x) x \int_{0}^{\ell}dx' \partial_{x'}
f(x')&\nonumber\\
	\approx \Theta(x) x \ell,
\end{align}
and one finds, with $f(x=0)=0$,
\begin{align}\label{eq:longrange}
	\chi_{\sigma} \approx \chi_{\sigma}^{(0)} - e \sigma B_p' \ell \left[ \Theta(x) - x_1/\ell\right] x - i \frac{eE_p'\ell}{v_F} \left[ \Theta(x) - x_0/\ell \right] x.
\end{align}
The derived state is bounded if $e^{\chi_{\sigma}(x)} \rightarrow 0$ for $x \rightarrow \pm \infty$ and this implies that
\begin{align}
	\sigma e B_p' \ell > \sigma e B_p' x_1 > 0,
	\end{align}
thus (i) $\sigma = {\rm sign}(B'_p \ell) = {\rm sign}(v_F \ell)$ and, (ii) $1 > x_1/\ell > 0$. These conditions are obtained for a non-uniform potential and are similar to what we have derived for the uniform magnetic field. We find respectively: (i) the existence of a unique $n = 0$ mode with iso-spin polarization (\ref{eq:polaf}) and (ii) a condition for localized states, as in Eq.~(\ref{eq:meanpost}). In that respect we find the same description of the $n = 0$ Fermi arc as for the strictly linear interface potential discussed in the previous sections such that this surface state is indeed topologically stable with respect to potential fluctuations. 

Moreover we find that the long-range behavior is of characteristic length $\Delta x_{L} = 1/e B_p' \ell$ while in the short-range regime we have found $\Delta x_{S} = \ell_S = 1/\sqrt{eB_p'}$. The two length scales are
represented in Fig.~\ref{fig:randominterface}. Within our linear interpolation, we have obtained the characteristic Gaussian wave functions of Landau quantization in the interface, and $\Delta x_S=\ell_S$ represents precisely their average width. On the contrary, we obtain an exponential decay of the surface states in the regions $x<0$ and $x>\ell$, as one expects from alternative treatments of sharp interfaces [\onlinecite{wsmsstheo1,wsmsstheo2, wsmsstheo3,wsmsstheo4}]. Because $B_p'\sim 1/\ell$, this long-range behavior is then, also in the present model, independent of the surface width $\ell$, as one may have expected. 
The following ratio
\begin{align}
	r = \Delta x_{S}/\ell = 1/\sqrt{eB_p\ell^2} = \sqrt{ v_F/(\Delta + \Delta')\ell},
\end{align}
quantifies the dominance of the long-range over the short-range regime. The use of boundary condition is valid in the long-range limit, which is defined by $r \gg 1$ and in the best case scenario ($\Delta' = 0$) one needs $\ell \ll 1/\Delta k \approx 1$nm.

In this section we have derived the Aharonov-Casher argument for a generic surface Hamiltonian and we have used it in order to argue the stability of the $n = 0$ Fermi arc of type-I and type-II WSM. This was also the opportunity to discuss the range of validity of boundary conditions. In the following we shall discuss the experimental relevance of all these results.

\section{Discussion and conclusion}
\label{sec:discussion}

In the present paper, we have investigated the surface states, i.e. the Fermi arcs, of Weyl semimetals within a model of ``soft'' interfaces between the semimetallic phase and an insulating phase. The main ingredient in our model is a set of parameters that characterize the different phases and that change linearly through the interface $x\in [0,\ell]$. This yields an interface (or surface) Hamiltonian that displays pseudo-magnetic and, in the case of tilted Weyl cones, pseudo-electric fields. From a formal point of view, the pseudo-magnetic field can be treated within Landau quantization, and the surface states can thus be viewed as Landau bands arising from this quantization. While the massive surface states, which correspond to Landau-band indices $n\neq 0$, are shifted to high energies for abrupt surfaces ($\ell\rightarrow 0$) or very large insulating gaps ($\Delta' \gg \Delta$), the $n=0$ Landau band is special in that it survives even in these limits. Furthermore, it remains stable also in the presence of fluctuations in the interface, i.e. when the confinement potential is not linear as supposed within our basic model. This topological stability of the $n=0$ is proven here with the help of generalization to 3D of the Aharonov-Casher argument [\onlinecite{acasher}]. 

\begin{figure}[t]
    \centering
    \includegraphics[width=\columnwidth]{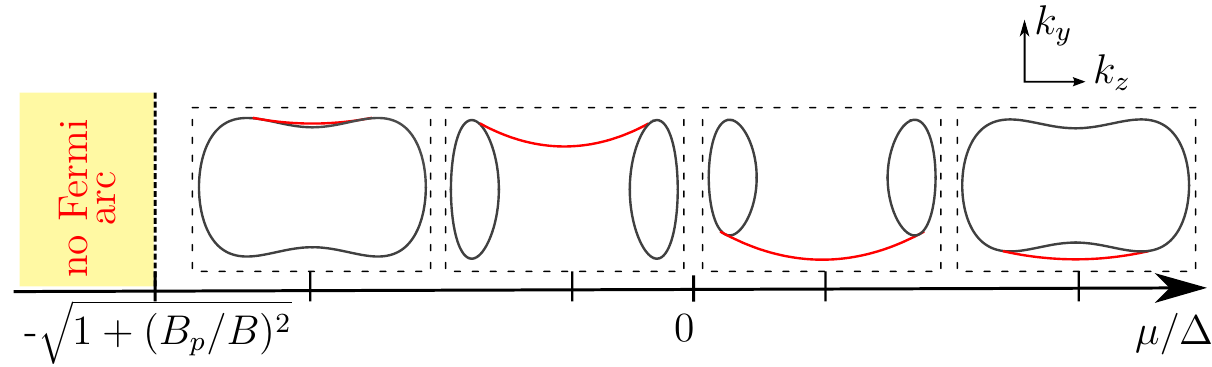}
    \caption{(Color online) Evolution of $B$-field deformed Fermi arcs when changing the chemical potential. The subfigures represent the section at $k_x = 0$ of the bulk Fermi surface of the WSM (in black) and the Fermi arc (in red) for a fixed external in-plane magnetic field ${\bf B} = B {\bf e}_z$, $B > 0$ and various values of the chemical potential $\mu$. The Fermi arc bends in the same orientation for positive and negative chemical potentials. For the chosen orientation of $B$, 
this bending can be such that it penetrates into the bulk Fermi surface, for values of the chemical potential below the Lifshitz transition, $\mu < -\sqrt{1+(B_p/B)^2}\Delta$.}
    \label{fig:magneticmu}
\end{figure}

The intersection between the $n=0$ topological surface state and the Fermi level yields the Fermi arcs characteristic of Weyl semimetals. More saliently, our model of a linear confinement at the interface and its treatment within Landau quantization allows us to understand in more detail the form and the manipulation of the Fermi arcs connecting the Weyl nodes. Indeed, the Fermi arc is a straight line if the line connecting the two Weyl nodes is simply shortened in the interface within a merging scenario. If, however, the line connecting the nodes is also rotated over the interface before the Weyl-point merging, the Fermi arcs are parabolically deformed as shown in Fig.~\ref{fig:Barcs}. We have shown that this parabolic deformation can also be achieved by the application of a magnetic field that we have chosen to be oriented in the $z$-direction connecting the Weyl nodes.
 Interestingly, the Fermi-arc survives even when the chemical potential $|\mu| > \Delta$, above the saddle point $S$ (see Fig.~\ref{fig:interface_general}). The corresponding Fermi surfaces are displayed on the rightest side of Fig.~\ref{fig:magneticmu} in the case $B > 0$ for positive values of $\mu$. Notice that the Fermi arc is always bent in the same direction for any chemical potential. Below a certain value of the chemical potential, the Fermi arc can be absorbed in the bulk Fermi sea and disappear, as sketched on the leftest side of Fig.~\ref{fig:magneticmu}.

Notice, however, that the possibility to manipulate the Fermi arcs with the help of a magnetic field is limited by the effective interface size $\ell_S=\sqrt{\ell/\Delta k}$, where $\Delta k$ is the reciprocal-space distance between the Weyl nodes. This size must be of the order of the magnetic length $\ell_B\simeq 25$ nm$/\sqrt{B{\rm [T]}}$, i.e. for an effective interface size in the 10 nm range, magnetic fields of the order of 10 T are required. We expect nevertheless a prominent effect of the magnetic field on the density of states of the Fermi arcs. Since the magnetic field can shorten or lengthen the Fermi arcs, the density of states is decreased or increased, respectively [see Eq.~(\ref{eq:DOS2})]. We expect this variation of the density of states $g_B(\mu)$ to influence the conductivity, $\sigma\propto g_B(\mu)$, such that one may obtain both a negative and a positive magneto-conductivity from the carriers in the Fermi arcs. Moreover, the magnetic field directly affects the group velocities $v_g$ of the electrons in the Fermi arcs and thus the conductivity, $\sigma\propto v_{g}^2$. As these effects are relevant in the limit of a magnetic length $\ell_B$ comparable to the interface length $\ell$, they can be used to experimentally evaluate $\ell$.

We have furthermore investigated the influence of the Weyl-cone tilt on the Fermi arcs. As shown in previous studies [\onlinecite{haoki1,sari,kawara}] moderate tilts can be absorbed by a Lorentz boost, preserving the Fermi arc structure. This is always the case for type-I WSM. However, in the case of type-II WSM, the effective tilt can be treated with the help of the Lorentz boosts only for some surface-orientations. Notice, that in the case of a finite sized insulating gap $\Delta' > 0$, if the tilt $|{\bf t}| > 1$ (type-II WSM), we can obtain a broken Fermi arc at any chemical potential. In any case the resulting $n = 0$ Fermi arc remains stable against fluctuations of the surface potential. We prove this fact using a generalized Aharonov-Casher argument.
\subsection*{Acknowledgements}
The authors thank A. Inhofer, B. Pla\c cais and D. Carpentier for fruitful discussions.

\appendix
\begin{widetext}
\section{Two-band Hamiltonian of Weyl nodes close to the merging transition}\label{ap:fourbandreduction}
We consider the following four band model of two coupled Weyl nodes with opposite chirality and opposite tilts $\mathbf{t}$
\begin{align}
	\hat{H}_{\rm c, 4} =&
	 \left( 
	\begin{array}{cc}
		v_F \mathbf{k} \cdot \left(\hat{\sigma} + \mathbf{t}\mathbbm{1}\right) + v_F \Delta k_0/2 \hat{\sigma_x} & \kappa \mathbbm{1} \\
		\kappa \mathbbm{1} & -v_F \mathbf{k} \cdot \left(\hat{\sigma} + \mathbf{t}\mathbbm{1}\right) + v_F \Delta k_0/2 \hat{\sigma_x}
	\end{array}	 \right).
\end{align}
One can perform the following change of basis 
\begin{align}
	&\hat{H}_{\rm c, 4}' = \hat{U} \hat{H}_{\rm c, 4} \hat{U}^{\dagger}\\
	&= 
	\left( 
	\begin{array}{cccc}
		-\left( \kappa + \frac{v_F \Delta k_0}{2} \right) & v_F( k_x - \mathbf{t} \cdot \mathbf{k} ) & 0 & v_F(k_z - ik_y)\\
		v_F( k_x - \mathbf{t} \cdot \mathbf{k} ) & \left( \kappa - \frac{v_F \Delta k_0}{2} \right) & v_F (k_z - i k_y)  & 0\\
		0 & v_F(k_z + ik_y) & -\left( \kappa - \frac{v_F \Delta k_0}{2} \right) &  -v_F( k_x + \mathbf{t} \cdot \mathbf{k} )\\
		v_F(k_z + ik_y) & 0 & -v_F( k_x + \mathbf{t} \cdot \mathbf{k} ) & \left( \kappa + \frac{v_F \Delta k_0}{2} \right)
	\end{array}
	\right)
\end{align}
\end{widetext}
with the new basis $\left( \Psi_{++}, \Psi_{+-}, \Psi_{-+}, \Psi_{--} \right)$ defined such as
\begin{align}
	\hat{U} = \frac{1}{2}\left(
		\begin{array}{cccc}
			 1 & -1 & -1 & 1 \\
			 -1 & 1 & -1 & 1 \\
			 -1 & -1 & 1 & 1 \\
			 1 & 1 & 1 & 1 \\
		\end{array}
	\right).
\end{align}
In the limit $|\kappa + v_F \Delta k_0/2| \gg |\kappa - v_F \Delta k_0/2|$, one can project (see Fig.~\ref{fig:band4full}) the Hamiltonian on the reduced $\left( \Psi_{+-}, \Psi_{-+} \right)$ basis
\begin{align}
	\hat{H}_{\rm c, 4}'\left(
	\begin{array}{c}
		\Psi_{++}\\
		\Psi_{+-}\\
		\Psi_{-+}\\
		\Psi_{--}
	\end{array}
	\right)
	\approx
	E
	\left(
	\begin{array}{c}
		0\\
		\Psi_{+-}\\
		\Psi_{-+}\\
		0
	\end{array}
	\right)
\end{align} 
where we neglect the terms of the form $E \Psi_{++}$ and $E \Psi_{--}$ since these are products of energy $E \approx 0$ and a small component $\Psi_{++}$ or $\Psi_{--}$. We develop the effective Hamiltonian up to first order in $k_y$ and $k_z$ and up to the second order in $k_x$ in order to describe the shift of the Weyl nodes along the $x-$axis. We then obtain the following Schr\"odinger equation
\begin{align}
	\hat{H}^{\rm eff}(\mathbf{k}) 
	\left(
	\begin{array}{c}
		\Psi_{+-}\\
		\Psi_{-+}
	\end{array}
	\right)
	=
	E
	\left(
	\begin{array}{c}
		\Psi_{+-}\\
		\Psi_{-+}
	\end{array}
	\right)
\end{align}
in terms of the effective two-band Hamiltonian
\begin{align}
	\hat{H}^{\rm eff}(\mathbf{k}) = &\frac{t_x ({\Delta k}/2)^2 - k_x \mathbf{t}\cdot\mathbf{k}}{2 m_0}\nonumber\\
	 &+  
	\left(
		\begin{array}{cc}
			 \frac{k_x^2 - ({\Delta} k/2)^2}{2m_1} & v_F(k_z - ik_y)\\
			 v_F(k_z + ik_y) & -\frac{k_x^2 - ({\Delta} k/2)^2}{2m_1}
		\end{array}
	\right),
\end{align}
where $m_0 = (\kappa + v_F \Delta k_0/2)/4v_F^2$, $m_1 = 2 m_0/(1+t_x^2)$ and ${\Delta}k = \Delta k_0\sqrt{1 - (2\kappa/v_F \Delta k_0)^2}/\sqrt{1+t_x^2}$. We have shifted the spectrum by $\Delta E = t_x \bar{\Delta k}^2/8 m_0$ to have cones at $E = 0$.

We can then adapt notations by permuting $k_x \leftrightarrow k_z$ and we obtain the following effective Hamiltonian
\begin{align}
	\hat{H}_{eff} &= t_z\left(\frac{k_z^2}{2 m} - \Delta\right) + \frac{v_F k_z}{\Delta k} (t_x k_x + t_y k_y)\\
	&+
	\left(
	\begin{array}{cc}
		\frac{k_z^2}{2 m} - \Delta & v_F (k_x - ik_y)\\
		v_F(k_x + i k_y) & -\left(\frac{k_z^2}{2 m} - \Delta\right)
	\end{array}
	\right)
\end{align} 
where $\Delta k = \sqrt{2 m \Delta}$.
\begin{figure}
    \centering
    \includegraphics[width=\columnwidth]{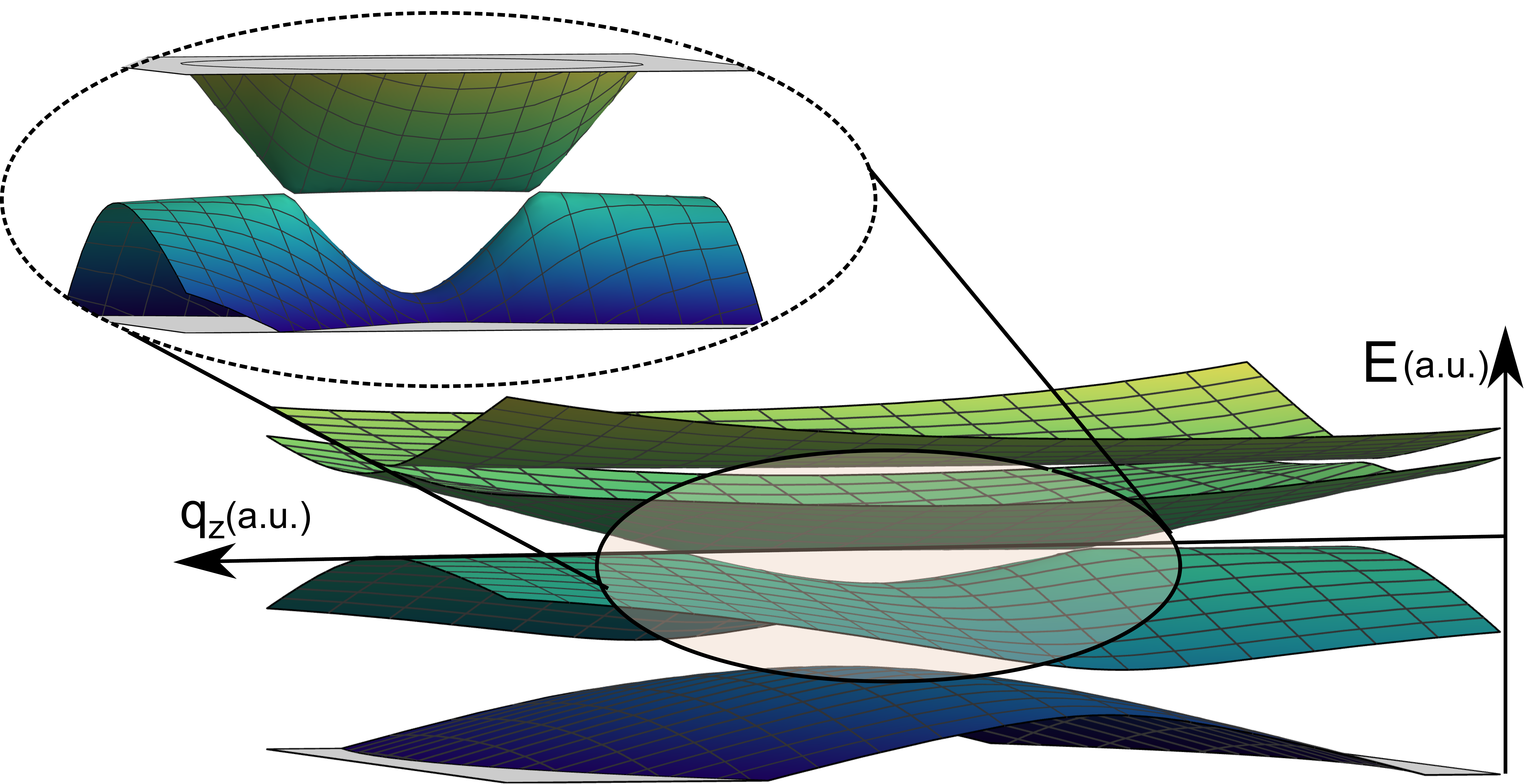}
    \caption{Sketch of the four band model of two coupled Weyl cones. The inset shows the low energy projection we perform in order to describe Weyl cones near fusion.}
    \label{fig:band4full}
\end{figure}

\section{On hyperbolic transformations}\label{ap:hyperbol}
In this article we refer to hyperbolic transformations which are known in relativistic quantum mechanics [\onlinecite{wignerot}] but which have rarely been used in condensed matter [\onlinecite{lukose}]. In this appendix we describe their representation on spinors and their relation to unitary transformation.
\subsection{Matrix representation} 
We introduce a transformation described by a generator $\hat{\Gamma}$ such that for a couple of two matrices $\hat{A}$ and $\hat{B}$ one has
\begin{align}
	\left\{
	\begin{array}{l}
	e^{\theta\hat{\Gamma}} \hat{A} e^{\theta \hat{\Gamma}} = \cosh(\theta) \hat{A} + \sinh(\theta) \hat{B},\\
	e^{\theta\hat{\Gamma}} \hat{B} e^{\theta \hat{\Gamma}} = \sinh(\theta) \hat{A} + \cosh(\theta) \hat{B}.
	\end{array}
	\right.
	\label{htr}
\end{align}
Such kind of transformation allows for the following transformation
\begin{align}
	e^{\theta\hat{\Gamma}} \left(\omega \hat{A} + v \hat{B} \right)  e^{\theta \hat{\Gamma}} &= \left[\cosh(\theta) \omega  + \sinh(\theta)v \right]\hat{A} \\&\nonumber+ \left[ \cosh(\theta) v  + \sinh(\theta)\omega\right] \hat{B},
\end{align}
and one can find a frame of reference where the dependence on matrix $\hat{B}$  vanishes if $\tanh(\theta) =  - v/\omega$, i.e. for $|v/\omega| < 1$. Similarly, if $\tanh(\theta) =  - \omega/v$, with $|\omega/v| < 1$, there exists a frame of references with no dependence on the matrix $\hat{A}$.
These two limits are similar to the space- and time- like limits of special relativity and are related to the sign of $v^2 - \omega^2$ which is a conserved quantity under hyperbolic transformations. We sketch the idea behind the two limits in Fig.~\ref{fig:minkowsky}.

\begin{figure}
    \centering 
    \includegraphics[width=0.5\columnwidth]{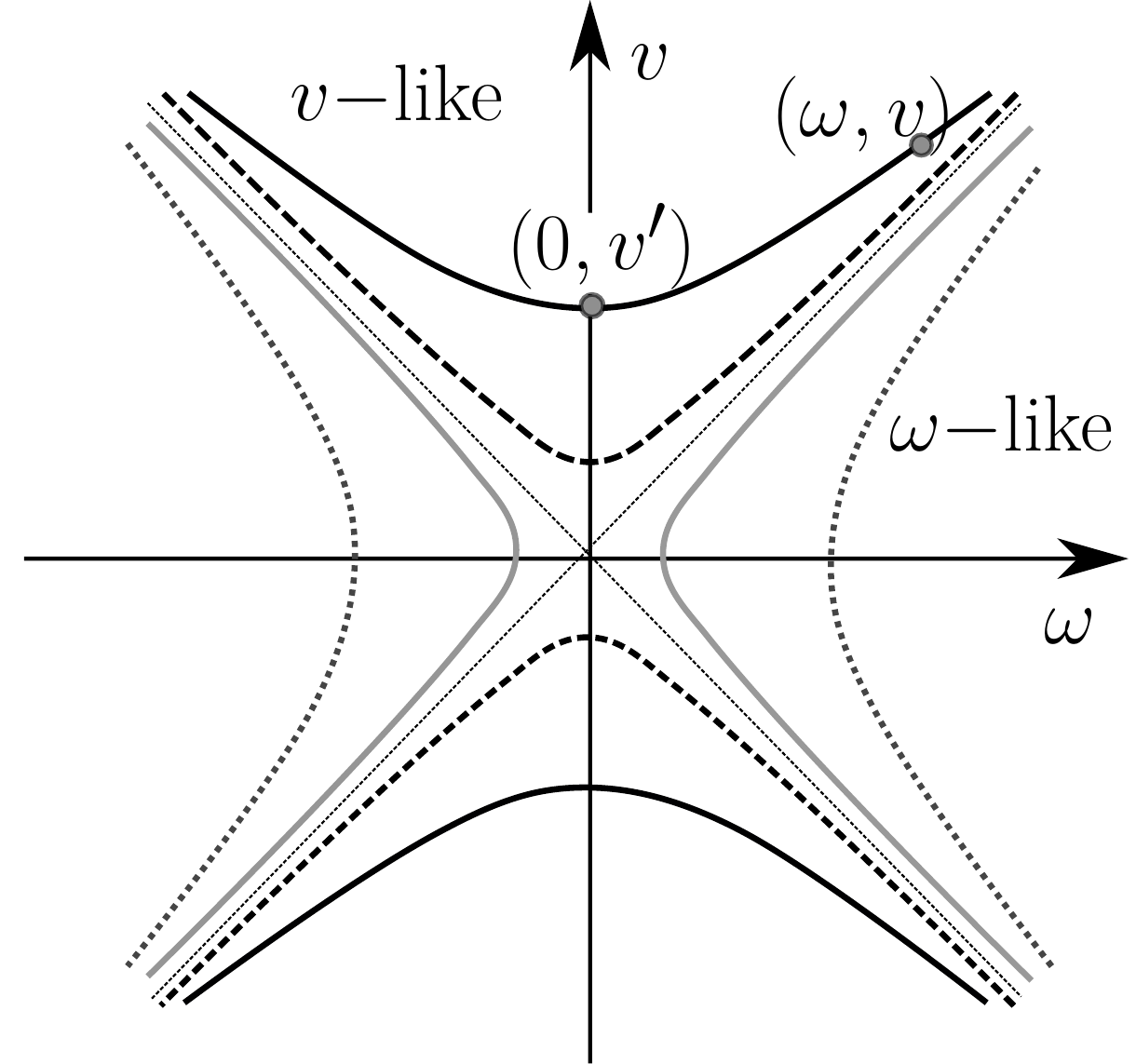}
    \caption{The set of parameters ($\omega$,$v$) can be represented as a 2D-plane on which the hyperbolic transformation link any set of parameters to any other on hyperbolic trajectories. One can get rid of one or the other parameter depending on their relative amplitude, this leads to the $\omega-$like and the $v-$like limits, similarly to the Minkowski space-time diagram of special relativity.}
    \label{fig:minkowsky}
\end{figure}

We now express the properties of the generator $\hat{\Gamma}$ by differentiating Eq.~(\ref{htr}) with respect to $\theta$. We then obtain the following anti-commutation rules for the generator $\hat{\Gamma}$
\begin{align}
	\left\{
	\begin{array}{l}
	\{ \hat{\Gamma} , \hat{A} \} = \hat{B},\\
	\{ \hat{\Gamma} , \hat{B} \} = \hat{A}.
	\end{array}
	\right.
	\label{antico0}
\end{align}
In the previous section we have discussed the way to absorb any position-dependent term on $\hat{A} = \mathbbm{1}$ into a trace-less matrix $\hat{B}$.  In this case Eq.~(\ref{antico0}) becomes
\begin{align}
	\left\{
	\begin{array}{l}
	\hat{\Gamma} = \hat{B}/2,\\
	\hat{B}^2 = \mathbbm{1}.
	\end{array}
	\right.
	\label{B2}
\end{align}
We observe that this transformation necessitates $\hat{B}^2 = \mathbbm{1}$ so that $\hat{B}$ can have only two real eigenvalues and this limitates the matrices on which this method can be used.

\begin{figure}
    \centering
    \includegraphics[width=\columnwidth]{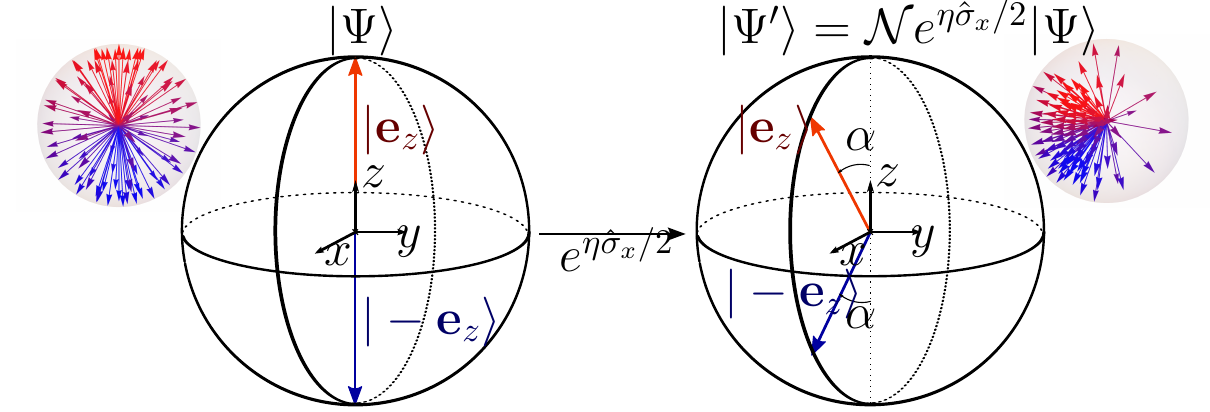}
    \caption{Representation of the states' deformation $|\pm{\bf e}_z\rangle$ on the Bloch sphere due to the hyperbolic transformation $e^{\eta \hat{\sigma}_x/2}$. In the insets, we picture the deformation of a homogeneous spin distribution over the Lorentz boost. The polarization effect is a combined effect of tilt and pseudo-magnetic field, as explained in the main text. Above a critical tilt, the polarization is strong enough to loose the hedgehog texture and the corresponding topological properties.}
    \label{fig:bloch_deformation}
\end{figure}

\subsection{Spinor representation}
In the main text we consider two-dimensional Hilbert spaces and perform hyperbolic transformations of the form $|\Psi\rangle \rightarrow |\Psi'\rangle = \mathcal{N} e^{\eta \hat{\sigma}_u /2} |\Psi \rangle$ with $\hat{\sigma}_u = \hat{\boldsymbol{\sigma}} \cdot {\bf u}$, ${\bf u}$ a unit vector. Any state $|{\bf n}\rangle$ can be written in terms of angles $(\theta, \phi)$ in the Bloch sphere
\begin{align}
	|{\bf n}\rangle = \left(
		\begin{array}{c}
			\cos(\theta/2)\\
			\sin(\theta/2)e^{i\phi}
		\end{array}
	\right),
\end{align}
and it fulfills $\langle {\bf n} | \hat{\sigma} | {\bf n} \rangle = {\bf n}$. Under the hyperbolic transformation one finds $|{\bf n'}\rangle = \mathcal{N} e^{\eta \hat{\sigma}_u /2} |{\bf n} \rangle$ with
\begin{align}
	{\bf n'} &= \frac{\langle {\bf n}' | \hat{\boldsymbol{\sigma}} | {\bf n}' \rangle}{\langle {\bf n}'| {\bf n}' \rangle} = \frac{\langle {\bf n} | e^{\eta \hat{\sigma}_u /2} \hat{\boldsymbol{\sigma}} e^{\eta \hat{\sigma}_u /2} | {\bf n} \rangle}{\langle {\bf n}| e^{\eta \hat{\sigma}_u } | {\bf n} \rangle}\\
	&= \frac{{\bf n}_{\perp} + \gamma ({\bf n}\cdot {\bf u} + \beta ){\bf u}}{\sqrt{({\bf n}_{\perp})^2 + \gamma^2({\bf n}\cdot {\bf u} + \beta)^2}},
\end{align}
with ${\bf n}_{\perp}$ is the components of ${\bf n}$ perpendicular to ${\bf u}$, $\gamma = \cosh(\eta)$ and $\beta = \tanh(\eta)$.

We consider the simplified case where ${\bf n}\cdot {\bf u} = 0$ and ${\bf n} = {\bf n}_{\perp}$, such that
\begin{align}
	{\bf n'} = \frac{\bf n}{\gamma} + \beta {\bf u},
\end{align}
and one observes that the Lorentz boost does not correspond to a simple rotation but a translation on the Bloch sphere. We express some useful results
\begin{align}
	&\mathcal{N} e^{-\eta_1 \hat{\sigma}_x /2} |\pm{\bf e}_z \rangle = e^{\pm i \theta_1 \hat{\sigma}_y/2} |\pm{\bf e}_z\rangle\\
	&\mathcal{N} e^{-\eta_2 \hat{\sigma}_y /2} e^{-\eta_1 \hat{\sigma}_x /2} |\pm{\bf e}_z \rangle = e^{\pm i \theta_s \hat{\sigma}_s/2} |\pm{\bf e}_z\rangle
\end{align}
with $\tan(\theta_1/2) = \sinh(\eta_1)$, $\tan(\theta_s/2) = \sqrt{\sinh^2(\eta_1)+\cosh^2(\eta_1)\sinh^2(\eta_2)}$ and $\hat{\sigma}_s = \hat{\boldsymbol{\sigma}}\cdot {\bf e}_s$, ${\bf e}_s = (-\sinh(\eta_2) {\bf e}_x + \tanh(\eta_1) {\bf e}_y)/\sqrt{\tanh(\eta_1)^2 + \sinh(\eta_2)^2}$.

\subsection{The Thomas-Wigner rotation}
\label{ap:wignerot}
We consider two Lorentz boosts $\hat{L}_1$ and $\hat{L}_2$ defined by $\hat{L}_i = e^{\eta_i \hat{\sigma}\cdot \mathbf{n}_i/2}$ where $\mathbf{n}_i^2 = 1$, $i \in \{1,2\}$. These operators are Hermitian, \emph{i.e.} $\hat{L}_i^{\dagger} = \hat{L}_i$, but their product is not
\begin{align}
	\hat{L}_1 \hat{L}_2 &= {\rm ch_1} {\rm ch_2} \left( \mathbbm{1} + {\rm th_1} \hat{\sigma}\cdot \mathbf{n}_1 \right)\left( \mathbbm{1} + {\rm th_2} \hat{\sigma}\cdot \mathbf{n}_2 \right)\\
	&= {\rm ch_1} {\rm ch_2} \left[ \left(1 + {\rm th_1} {\rm th_2} \mathbf{n}_1 \cdot \mathbf{n}_2 \right) \mathbbm{1}\right.\nonumber\\
	&~~~~~~~~~\left.+ \left( {\rm th_1} \mathbf{n}_1 + {\rm th_2} \mathbf{n}_2 + i {\rm th_1} {\rm th_2} \mathbf{n}_1\times\mathbf{n}_2 \right)\cdot \hat{\sigma} \right]
\end{align}
where we use that $\hat{\sigma}_i \hat{\sigma}_j = i\varepsilon_{ijk} \hat{\sigma}_k + \delta_{ij} \mathbbm{1}$ and write $\cosh(\eta_i/2) = {\rm ch_i}$ and $\tanh(\eta_i/2) = {\rm th_i}$. Since the product of two Lorentz boosts is not hermitian, it cannot be a Lorentz boost but is actually the combination of a rotation, $\hat{R}_{\theta} = e^{i \theta \hat{\sigma}\cdot \mathbf{u}/2}$, and a Lorentz boost, $\hat{L}_3 = e^{\eta_3 \hat{\sigma}\cdot \mathbf{n}_3/2}$, where $\mathbf{u}^2 = 1$ and $\mathbf{n}_3^2 = 1$. One can compute $\hat{L}_3\hat{R}_{\theta}$
\begin{align}
	\hat{L}_3\hat{R}_{\theta} &= {\rm ch_3}c \left(  \mathbbm{1} + {\rm th_3} \hat{\sigma} \cdot \mathbf{n}_3 \right)\left(  \mathbbm{1} + it \hat{\sigma} \cdot \mathbf{u} \right)\\
	&= {\rm ch_3}c \left[ \left( 1 + i t{\rm th_3} \mathbf{u}\cdot \mathbf{n}_3 \right)\mathbbm{1}\right.\nonumber\\
	&~~~~~~~~~\left.+ \left( {\rm th_3} \mathbf{n}_3 + it \mathbf{u} + t {\rm th_3} \mathbf{u}\times\mathbf{n}_3 \right) \cdot \hat{\sigma} \right]
\end{align}
wit $\cos(\theta/2) = c$ and $\tan(\theta/2) = t$. We then identify $\hat{L}_1\hat{L}_2 = \hat{L}_3 \hat{R}_{\theta}$ and apply the projection operator $\hat{P}_{i}\bullet = {\rm Tr}\left(\hat{\sigma}_i \bullet \right)$ for $i = 0,1,2,3$ on this equation. We identify the real and imaginary parts and find
\begin{align}\label{eq:apallwigeq}
	\left\{
	\begin{array}{l}
	{\rm ch_1} {\rm ch_2}\left( 1 + {\rm th_1} {\rm th_2} \mathbf{n}_1\cdot\mathbf{n}_2 \right) = {\rm ch_3} c,\\
	\mathbf{u}\cdot \mathbf{n}_3 = 0,\\
	{\rm ch_1} {\rm ch_2} \left( {\rm th_1} \mathbf{n}_1 + {\rm th_2} \mathbf{n}_2 \right) =  {\rm ch_3} {\rm th_3} c \left(  \mathbf{n}_3 + t \mathbf{u}\times\mathbf{n}_3 \right),\\
	{\rm ch_1} {\rm ch_2} {\rm th_1} {\rm th_2} \mathbf{n}_1 \times \mathbf{n}_2 = {\rm ch_3} c t \mathbf{u}.
	\end{array}\right.
\end{align}
We combine these equations and find
\begin{align}\label{eq:apwignerot}
	\left\{
	\begin{array}{l}
	t \mathbf{u} = \frac{{\rm th_1} {\rm th_2}}{1 + {\rm th_1} {\rm th_2} \mathbf{n}_1 \cdot \mathbf{n}_2} \mathbf{n}_1\times\mathbf{n}_2,\\~~\\
	{\rm th_3} \mathbf{n}_3 = \frac{\left( 1 + {\rm th_2}^2 + 2{\rm th_1} {\rm th_2} \mathbf{n}_1 \cdot \mathbf{n}_2 \right)  {\rm th_1} \mathbf{n}_1 + \left( 1 - {\rm th_1}^2 \right)  {\rm th_2} \mathbf{n}_2}{1 + ({\rm th_1} {\rm th_2})^2 + 2{\rm th_1} {\rm th_2} \mathbf{n}_1\cdot\mathbf{n}_2},
	\end{array}
	\right.
\end{align}
which express the corresponding rotation angle $\theta$ and boost $\eta_3$. In the limit $\mathbf{n}_1 \equiv \mathbf{n}_2$ we find no rotation, $t = 0$, and
\begin{align}
	\tanh(\eta_3/2) &= \frac{\tanh(\eta_1/2) + \tanh(\eta_2/2)}{1+\tanh(\eta_1/2)\tanh(\eta_2/2)}\\
	&= \tanh((\eta_1 + \eta_2)/2)
\end{align}
which implies $\eta_3 = \eta_1 + \eta_2$ as expected from geometrical arguments in special relativity. In general, one finds that two consecutive boosts involve a rotation, the \emph{Thomas-Wigner rotation}. For example, in (\ref{eq:apdefhyp}) one has $\mathbf{n}_1 = \mathbf{e}_x$ and $\mathbf{n}_2 = \mathbf{e}_y$ and one observes a rotation along $\mathbf{e}_x\times\mathbf{e}_y = \mathbf{e}_z$ with an angle $\theta$ such that
\begin{align}
	\tan(\theta/2) = \tanh(\eta_1/2)\tanh(\eta_2/2).
\end{align}

\end{document}